\documentclass{article}

\usepackage{arxiv}

\usepackage[utf8]{inputenc} 
\usepackage[T1]{fontenc}    
\usepackage{hyperref}       
\usepackage{url}            
\usepackage{booktabs}       
\usepackage{amsfonts}       
\usepackage{nicefrac}       
\usepackage{microtype}      
\usepackage{lipsum}		
\usepackage{graphicx}
\usepackage{natbib}
\usepackage{doi}

\usepackage{amsthm,amsmath} 
\usepackage{mathrsfs} 
\usepackage{amsmath}
\usepackage{amssymb}
\usepackage{latexsym}
\usepackage{bm}
\usepackage{color}
\usepackage{url} 
\newcommand{\choosefont}[1]{\fontfamily{#1}\selectfont}

\usepackage{mathtools} 
\usepackage{algorithmic} 
\usepackage{algorithm}
\usepackage{setspace}
\usepackage{multirow}
\usepackage{booktabs}

\theoremstyle{plain}
\newtheorem{thm}{Theorem}
\newtheorem*{thm*}{Theorem}

\theoremstyle{definition}

\newtheorem{lem}[thm]{Lemma} 
\newtheorem{prop}{Proposition} 

\title{Wilcoxon-type Multivariate Cluster Elastic Net}

\author{ \href{https://orcid.org/0000-0001-9944-9569}{\includegraphics[scale=0.06]{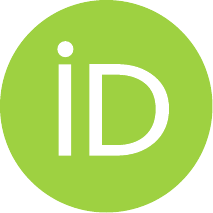}\hspace{1mm}Mayu Hiraishi} \\ 
	Graduate School of Culture and Information Science\\
	Doshisha University\\
	Kyoto, Japan\\
	\texttt{mayumonnn@gmail.com} \\
	\And
	\href{https://orcid.org/0000-0001-9621-5871}{\includegraphics[scale=0.06]{orcid.pdf}\hspace{1mm}Kensuke Tanioka} \\
	Department of Biomedical Sciences and Informatics\\
	Doshisha University\\
	Kyoto, Japan\\
	\texttt{ktanioka@mail.doshisha.ac.jp} \\
	\And
	\href{https://orcid.org/0000-0002-1408-2655}{\includegraphics[scale=0.06]{orcid.pdf}\hspace{1mm}Hiroshi Yadohisa} \\
	Department of Culture and Information Science\\
	Doshisha University\\
	Kyoto, Japan\\
	\texttt{hyadohis@mail.doshisha.ac.jp} \\
}

\renewcommand{\shorttitle}{\textit{arXiv} Template}

\hypersetup{
pdftitle={A template for the arxiv style},
pdfsubject={q-bio.NC, q-bio.QM},
pdfauthor={David S.~Hippocampus, Elias D.~Striatum},
pdfkeywords={First keyword, Second keyword, More},
}

\begin{document}
\maketitle

\begin{abstract}
We propose a method for high dimensional multivariate regression that is robust to heavy-tailed distributions or outliers, while preserving estimation accuracy in normal random error distributions. We extend the Wilcoxon-type regression to a multivariate regression model as a tuning-free approach to robustness. Furthermore, the proposed method regularizes the $L_1$ and $L_2$ terms of the clustering based on $k$-means, which is extended from the multivariate cluster elastic net. The estimation of the regression coefficient and variable selection are produced simultaneously. Moreover, considering the relationship among the correlation of response variables through the clustering is expected to improve the estimation performance. The numerical simulation demonstrates that our proposed method overperforms the multivariate cluster method and other multiple regression methods in the case of heavy-tailed error distribution and outliers. The proposed method also indicates stability in normal error distribution. Finally, we confirm the efficacy of our proposed method using gene data.
\end{abstract}

\keywords{MM algorithm \and multivariate regression \and robust statistics}

\section{Introduction}
Numerous studies on multivariate regression analysis have proposed methods to demonstrate the efficiency of the estimation. As part of this framework, it is expected that the accuracy of the estimation is enhanced by considering correlations among the objective variables. 
\citet{breiman1997} proposed the Curds and Whey method, which predicts multivariate response variables with an optimal linear combination of least squared predictors.
\citet{MRCE} proposed a multivariate regression with covariance estimation, which estimates regression coefficients and inverse error covariance simultaneously. 
To improve estimation, multivariate regression has been extended to include a sparse regularization term for  high-dimensional data, such as genetic data \citep{peng2010regularized,kim2012tree,chen2017modeling} . 
These methods consider the correlations of response variables.
Methods for reducing the dimension of the response have also been introduced by \citet{cook2010envelope}, \citet{cook2015foundations}, and \citet{sun2015sprem}. 
By contrast, the multivariate cluster elastic net (MCEN) is proposed by \citet{MCEN} as a multivariate regression that considers the correlation of the response variables without a prior information. 
This method includes an $L_1$ penalty \citep{lasso} for the variable selection and $L_2$ penalty for identifying clusters based on $k$-means \citep{forgy1965} associated with the response variables. This method differs from the previously mentioned methods in that the estimation of the regression coefficients and grouping of the response variables is conducted simultaneously. 
For data with a small number of samples and a large number of explanatory variables, such as genetic data, the lasso regularization term helps to estimate a sparse coefficient matrix. The multivariate cluster elastic net can be also applied to high-dimensional data with multiple response variables. 
Another characteristic of the multivariate cluster elastic net is the grouping of the fitted values of the response variables by the $k$-means clustering term. 
With this term, the multivariate cluster elastic net can improve estimation accuracy by considering correlations between the response variables. \\

Meanwhile, in a regression model, the error distributions are assumed to be Gaussian. 
However, when the errors follow a heavy-tailed distribution or contain outliers, this may affect the estimation results (e.g., \citep{huber2011robust}).
Various robust regression methods have been proposed to solve this problem \citep{fan2017estimation, loh2017statistical, sun2019adaptive, Lozano2016, Wang2013, AvellaMedina2018RobustAC, Prasad2020RobustEV, bellobi2011, bradic2011penalized, wang2012quantile, wang2013l1,Fan2014ADAPTIVERV, sun2012scaled}.
Huber regression reduces the impact of outliers by placing an adjustment parameter \citep{Huber1964RobustEO}.
Instead of least squares, the lad-lasso \citep{wang2007robust} sets the least absolute deviation as its loss function. \citet{zou2008composite} introduced the composite quantile regression. 
The rank-based estimate with Wilcoxon scores \citep{jureckova1971nonparametric, jaeckel, hettmansperger1977robust} is also known to be more efficient than least squared procedures when the data deviates from a normal distribution. This is because it is robust against a response containing outliers \citep{hettmansperge1998}. 
This Wilcoxon estimate does not need a tuning parameter for robust estimation. 
\citet{wlasso} and \citet{wwscad} use a Wilcoxon estimate as their loss function with a regularization penalty for robust estimation in a high-dimensional setting. However, these methods remain in the framework of multiple regression.\\

We propose a Wilcoxon-type regression that is extended to a multivariate cluster elastic net. This is a high-dimensional regression with robustness to heavy-tailed error distribution or outliers containing random errors. We  call this new method the Wilcoxon-type multivariate cluster elastic net (WMCEN). 
The Wilcoxon-type loss function is expected to be robust to heavy-tailed error distribution or outliers. The proposed method is also suitable for a high-dimensional situation because the variable selection uses the $L_1$ regularization penalty. 
Moreover, the proposed method obtains more efficiency because the correlation of the response variables is considered, following a similar framework to that of the multivariate cluster elastic net. 
When updating the parameters of the proposed method, we derive the majorizing function of the updated formula based on the majorize-minimization (MM) algorithm \citep{MMtutorial}. With the majorizing function, an updated formula based on the squared error criterion can be calculated, which ensures that the updated formula can be solved easily. 
Wilcoxon-type regression in multivariate regression has been proposed by \citet{multiWil}. This Wilcoxon-type regression is extended to deal with multiple response variables based on $L_2$ norm, and the theoretical properties are guaranteed. In this study, we adopted a different extension to multivariate outcomes in order to enhance the familiarity in the updated formula of the proposed method with that of the original MCEN. \\

Section $2$ shows the objective function of our proposed method and explains the derivation of the majorizing function. After describing the algorithm and the updated formula of the proposed method, Section $3$ shows the efficiency of the proposed method through numerical simulation. Section $4$ describes the application of the proposed method to real genetic data related to breast cancer. Section $5$ concludes the paper with directions for future studies.

\section{Wilcoxon-type Multivariate Cluster Elastic Net
(WMCEN)}
\label{sec2}

This section explains the optimization problem of our proposed method and introduces the majorizing function corresponding to the new method.
We first explain the objective function of Wilcoxon-type regression, which is the basis of the extension of our proposed method. 
Let $y_i \ (i=1,2,\cdots, n)$ be response variables, where $n$ is the number of subjects. Let ${\bm{x}_i} \in \mathbb{R}^p \ (i=1,2,\cdots, n)$ be covariates in $p$ dimension, and ${\bm \beta} \in \mathbb{R}^p $ be unknown regression coefficients in $p$ dimension.
The Wilcoxon-type regression method for a single outcome is estimated by minimizing $\bm{\beta}$ as follows:
\begin{align}
\sum_{i<j}|e_i-e_j|
\label{dispersion}
\end{align}
where $e_i=y_i-\bm{x}_i^T\bm{\beta}, e_j=y_j-\bm{x}_j^T\bm{\beta} \; (i,j = 1, 2, \cdots, n)$. 
$e_i$ and $e_j$ are expressed in the form of a linear regression model, and Eq. (\ref{dispersion}) represents the absolute sum of the difference of the residuals between $i$ and $j$. This function in Eq. (\ref{dispersion}) is equivalent to Jaeckel's Wilcoxon-type dispersion function \citep{jaeckel}:
\begin{align}
\sqrt{12}\sum_{i=1}^n \left[ \frac{\bm{R}(y_i - \bm{x}_i^T \bm{\beta})}{n+1}-\frac{1}{2}\right](y_i - \bm{x}_i^T \bm{\beta})
\label{jaeckel}
\end{align}
where $\bm{R}(y_i - \bm{x}_i^T \bm{\beta})$ is the rank of $(y_i - \bm{x}_i^T \bm{\beta}) \; (i=1,2,\cdots,n)$ \citep{HM1978}. That is, Eq. (\ref{jaeckel}) can be regarded as a weighted regression.

\subsection{Optimization problem of Wilcoxon-type multivariate cluster elastic net}

Prior to describing the optimization problem of the proposed method, we introduce the multivariate regression model in this subsection. 
Let $\bm{Y}=(\bm{y}_1,\bm{y}_2,\cdots, \bm{y}_n)^T \in \mathbb{R}^{n\times q}$ be matrix of responses, ${\bm X}=(\bm{x}_1,\bm{x}_2,\cdots, \bm{x}_n)^T \in \mathbb{R}^{n\times p}$ be regression covariates, ${\bm B}=(\bm{\beta}_1,\bm{\beta}_2,\cdots, \bm{\beta}_q) \in \mathbb{R}^{p\times q}$ be unknown regression coefficients, and $\bm{E}=(\bm{e}^\dagger_1, \bm{e}^\dagger_2, \cdots, \bm{e}^\dagger_n)^T$ be i.i.d. random errors. The linear regression model is then represented as follows:
\begin{align*}
\bm{Y}=\bm{XB}+\bm{E}.
\end{align*}
We now explain the optimization problem of the WMCEN. To achieve this, we extend the Wilcoxon-type regression method to a multivariate regression method as follows: 
\begin{align}
L^* (\{{\bm \beta}_s\}_{s=1}^q)=
\sum_{s=1}^q \sum_{i<j}|e_{is} -e_{js}|  \rightarrow{\rm Min}
\label{MultiWil} 
\end{align}
where $e_{is}=y_{is}-\bm{x}_i^T\bm{\beta}_s, e_{js} = y_{js}-\bm{x}_j^T\bm{\beta}_s \; (i,j = 1, 2, \cdots, n; s = 1,2, \cdots, q)$.
Eq.(\ref{MultiWil}) represents the sum of the objective function of the Wilcoxon-type regression method for each response variable.
We extend the Wilcoxon-type regression to the multivariate cluster elastic net (MCEN), and formulate the optimization problem of WMCEN based on Eq.(\ref{MultiWil}) as follows:
\begin{align}
L(\{{\bm \beta}_s\}_{s=1}^q, {\bm{U}}, \{{\bm v}_\ell\}_{\ell=1}^k) = &
\sum_{s=1}^q \sum_{i<j}|e_{is} -e_{js}| 
+\lambda \sum^{q}_{s=1}\|{\bm \beta}_s\|_1 \nonumber\\ 
&+ \frac{\gamma}{2} \sum_{s=1}^q \sum_{\ell=1}^k u_{s\ell}\|{\bm X}{\bm \beta}_s - {\bm X}{\bm v}_\ell\|_F^2 \rightarrow{\rm Min}
\label{WMCEN1}
\end{align}
where $\lambda \; ( \lambda >0)$ and $\gamma \; (\gamma >0) $ are the tuning parameters. 
$\|\cdot\|_F$ is the Frobenius norm, $\|\cdot\|_1$ is the $L_1$ norm, and $\|\cdot\|_2$ is the $L_2$ norm. 
The second and third terms of Eq. (\ref{WMCEN1}) are the same regularization terms as those used in the multivariate cluster elastic net \citep{MCEN}.
${\bm U} = (u_{s\ell}) \ (s=1,2,\cdots, q; \ell = 1, 2, \cdots, k)$ is the indicator matrix, which represents the degree of belonging to cluster $\ell$ on the response variable $s$. 
When $u_{s\ell} = 1$, ${\bm X}{\bm \beta}_s$ belongs to the $\ell$th cluster, otherwise $u_{s\ell} = 0$.
$\bm{v}_\ell \; (\ell = 1, 2, \cdots, k$) is the partial regression coefficient for cluster centroid.
The second formula term is the lasso penalty \citep{lasso}, and 
the third term of Eq. (\ref{WMCEN1}) is the $k$-means clustering function.\\
This study derives the updated formula using the MM algorithm \citep{MMtutorial}.
To solve the problem of applying the MM algorithm to the lasso penalty \citep{HandL}, we substitute the perturbed version of the lasso penalty term for the traditional lasso penalty term \citep{MMlasso}.
\begin{align}
L^{\dag}(\{{\bm \beta}_s\}_{s=1}^q, {\bm{U}}, \{{\bm v}_\ell\}_{\ell=1}^k) 
= & 
\sum_{s=1}^q \sum_{i<j}|e_{is} -e_{js}| 
+\lambda\sum_{s=1}^q\sum_{h=1}^p\left(|\beta_{sh}|-\epsilon\log \left(1+\frac{|\beta_{sh}|}{\epsilon}\right)\right) \nonumber\\
& + \frac{\gamma}{2} \sum_{s=1}^q \sum_{\ell=1}^k u_{s\ell}\|{\bm X}{\bm \beta}_s - {\bm X}{\bm v}_\ell\|_F^2
\label{WMCEN}
\end{align}
where $\epsilon \; (\epsilon >0)$ is the hyper parameter of the lasso penalty.
When $\epsilon$ approaches $0$, the second term of Eq. (\ref{WMCEN}) converges to the second term of Eq. (\ref{WMCEN1}).
\\

\noindent
{\rm \bf{Remark 1}}\\
\citet{multiWil} has proposed Wilcoxon-type regression for multivariate regression. The model is formulated as
\begin{align}
D(\bm{B})=\sum\sum_{i<j}\|\bm{e}_i-\bm{e}_j\|_2
\label{zhou}
\end{align}
where $\bm{e}_i=\bm{y}_i-\bm{B}^T\bm{x}_i, \bm{e}_j = \bm{y}_j -\bm{B}^T\bm{x}_j \; (i,j = 1, 2, \cdots, n).$ Here, $\| \cdot \|_2$ represents $L_2$ norm. Eq. (\ref{zhou}) extends Wilcoxon-type regression to multivariate regression using $L_2$ norm. By contrast, the proposed method extends it to multivariate regression using the $L_1$ norm, as shown in Eq. (\ref{MultiWil}). 
By using $L_1$ norm, the proposed method can utilize the updated formula of the existing MCEN combining with the MM algorithm, as discussed in Section $2.3.$.\\

\noindent
{\rm \bf{Remark 2}}\\
Theoretical properties of the proposed method without regularization terms, that is Eq. (\ref{MultiWil}), are different from those of Wilcoxon-type regression using $L_2$ norm. For Wilcoxon-type regression based on $L_2$ norm, asymptotic properties of estimators are presented in \citet{multiWil}. On the other hand, multivariate Wilcoxon-type regression using $L_1$ norm without regularization terms can be guaranteed by existing model for single response \citep{chang1999high, heiler1988}. The parameter estimation problem of Eq. (\ref{MultiWil}) is equivalent to the following problem
\begin{align}
\sum_{i<j}|e_{is}-e_{js}|
\label{prop1}
\end{align}
by each $s \ (s=1, 2, \cdots, q)$. Therefore, the approximation theory in \citet{chang1999high}, and \citet{heiler1988} can be applied to multivariate Wilcoxon-type regression using $L_1$ norm without regularization problem.\\

\noindent
{\rm \bf{Remark 3}}\\
Next, the approximation theory of multivariate Wilcoxon-type regression using $L_1$ with penalty terms related to sparsity, that is the first and second terms of Eq. (\ref{WMCEN1}), is also shown in the same manner of {\rm Remark 2} \citep{wwscad, johnson2008rank}.

\subsection{Deriving the majorization function of WMCEN
}

In this subsection, we derive the majorizing function from Eq. (\ref{MultiWil}), as shown in Lemma \ref{lem1}, and the majorizing function of the perturbed lasso penalty is derived in Lemma \ref{lem2}. With Lemma \ref{lem1} and Lemma \ref{lem2}, we provide the majorizing function of the proposed method in Proposition \ref{propo1}.

We first derive the majorizing function of the first term of Eq. (\ref{WMCEN}). The objective function of WMCEN in Eq. (\ref{MultiWil}), that is the first term of Eq. (\ref{WMCEN}), can be transformed as follows:
%
\begin{align}
\sum_{s=1}^q \sum_{i<j}|e_{is} -e_{js}| 
&= \sum_{s=1}^q \sum_{i<j}|(y_{is}-\bm{x}_i^T\bm{\beta}_s) - (y_{js}-\bm{x}_j^T\bm{\beta}_s)| \nonumber\\
&= \sum_{s=1}^q \sum_{i<j}|(y_{is} -y_{js}) - (\bm{x}_{i} -\bm{x}_{j})^T\bm{\beta}_s| \nonumber\\
&= \sum_{s=1}^q \sum_{i<j}|g_{ijs} - \bm{r}_{ij}^T\bm{\beta}_s| 
\label{MM_1t}
\end{align}
where $g_{ijs} = y_{is} -y_{js}$, and $\bm{r}_{ij} = \bm{x}_{i} -\bm{x}_{j} \; (i,j=1,2,\cdots,n;\ i<j)$. $g_{ijs}$ represents the difference between the $i$ and $j$ of the $s$th response variable, and $\bm{r}_{ij}$ is the difference between the $i$ and $j$ of the covariate vectors. By forming them in Eq. (\ref{MM_1t}), the difference of the residuals can be expressed as a regression.\\

Here, we explain the majorizing function.
$\theta$ is set as the parameter of the interested real-valued objective function $f(\theta)$, and $\theta^{(t)}$ denotes the fixed value of parameter $\theta$.
In the algorithm, for example, $\theta^{(t)}$ represents the estimated $\theta$ at the $t$th step.
$g(\theta | \theta^{(t)})$ also denotes the function, and the $g(\theta | \theta^{(t)})$ can easily derive the updated formula when given $\theta^{(t)}$. The function $g(\theta | \theta^{(t)})$ is defined as the majorizing function of $f(\theta)$ at $\theta^{(t)}$ when the following two conditions are met:
\begin{align*}
g(\theta|\theta^{(t)}) & \geq f(\theta) \ {\rm for \ all} \ \theta,\nonumber\\
g(\theta^{(t)}|\theta^{(t)}) &= f(\theta^{(t)}).
\end{align*}
Next, we obtain the first term of the majorizing function from Eq. (\ref{MM_1t}).

\begin{lem}
Given $g_{ijs}, {\bm r}_{ij}$, and ${\bm \beta}_s^{(t)}$, for any ${\bm \beta}_s$, the following inequality holds:
\begin{align}
\sum_{s=1}^q \sum_{i<j}|g_{ijs}-{\bm r}_{ij}^T{\bm\beta}_s|
\leq \sum_{s=1}^q \sum_{i<j} \frac{1}{2}\frac{|g_{ijs}-{\bm r}_{ij}^T{\bm \beta}_s|^2}{|g_{ijs}-{\bm r}_{ij}^T{\bm \beta}_s^{(t)}|}
+\sum_{s=1}^q \sum_{i<j} \frac{1}{2}|g_{ijs}-{\bm r}_{ij}^T{\bm \beta}_s^{(t)}|.
\label{MM1dd}
\end{align}
where ${\bm \beta}_s^{(t)}$ is the $t$th fixed ${\bm \beta}_s$ in the algorithm. If ${\bm \beta}_s^{(t)} = {\bm \beta}_s$ for all $s  = 1, 2, \cdots, q$, the equality of Eq. (\ref{MM1dd}) holds.
\label{lem1}
\end{lem}

From {\rm Lemma \ref{lem1}}, the first term of right side in Eq. (\ref{WMCEN}) can be derived as a quadratic function. Subsequently, for any ${\bm \beta}_s$, this majorizing function can be described as follows:
\begin{align}
\sum_{s=1}^q \sum_{o=1}^{n(n-1)/2} w_{os}{|g_{os}-{\bm r}_o^T{\bm \beta}_s|^2}+w_{os}
\label{MM1st}
\end{align}
where $w_{os}=\frac{1}{2|g_{os}-{\bm r}_o^T{\bm \beta}_s^{(t)}|}$. Let ${\bm \beta}_s^{(t)}$ be a fixed $p$-dimensional vector, and the equality of Eq. (\ref{MM1st}) is established when ${\bm \beta}_s= {\bm \beta}_s^{(t)}$. Because Eq. (\ref{MM1st}) is a weighted multivariate regression, predicted values can be obtained using the conventional squared error criterion.\\

The second term of the majorizing function is then provided from the second term of Eq. (\ref{WMCEN}).

\begin{lem}
Given $\lambda \ (\lambda>0)$, $\epsilon \ (\epsilon>0)$, and fixed  $\beta_{sh}^{(t)}$, for any ${\bm \beta}_s$, the following inequality holds:%
\begin{align}
&\lambda\sum_{s=1}^q\sum_{h=1}^p\left(|\beta_{sh}|-\epsilon\log \left(1+\frac{|\beta_{sh}|}{\epsilon}\right)\right) \nonumber \\
& \leq \lambda \sum_{s=1}^q\sum_{h=1}^p \left( |{\beta}_{sh}^{(t)}|-\epsilon\log \left(1+\frac{|{\beta}_{sh}^{(t)}|}{\epsilon}\right)
+ \frac{({\beta}_{sh})^2-({\beta}_{sh}^{(t)})^2}{2(|{\beta}_{sh}^{(t)}|+\epsilon)}\right)
\label{lam}
\end{align}
where the equality of Eq. (\ref{lam}) holds only when $\beta_{sh} = \beta_{sh}^{(t)}$ for all $\beta_{sh}$. \\
\label{lem2}
\end{lem}

With {\rm Lemma \ref{lem2}}, we show the majorizing function can be derived from the perturbed version of the lasso penalty as the squared form.
The right-hand side of Eq. (\ref{lam}) can be expressed as
\begin{align}
&\lambda \sum_{s=1}^q\sum_{h=1}^p \left( |{\beta}_{sh}^{(t)}|-\epsilon\log \left(1+\frac{|{\beta}_{sh}^{(t)}|}{\epsilon}\right)
+ \frac{({\beta}_{sh})^2-({\beta}_{sh}^{(t)})^2}{2(|{\beta}_{sh}^{(t)}|+\epsilon)}\right) 
= \|{\bm \Psi}_s{\bm \beta}_s\|_F^2+C 
\label{MMpert}
\end{align}
where 
\begin{align*}
{\bm \Psi}_s = {\rm diag}\left(\frac{1}{\sqrt{2({\beta}_{s1}^{(t)}+\epsilon)}},
\frac{1}{\sqrt{2({\beta}_{s2}^{(t)}+\epsilon)}},\cdots,\frac{1}{\sqrt{2({\beta}_{sp}^{(t)}+\epsilon)}} \right),
\end{align*}
and $C$ represents constant values that are not relevant to $\beta_{sh}$. 
From Lemma \ref{lem1} and Lemma \ref{lem2}, the following proposition holds.


\begin{prop}\label{propo1}
Given ${\bm Y}, {\bm X}, \lambda \; (\lambda>0), \gamma \; (\gamma >0),$ and $\epsilon \; (\epsilon > 0 )$, the following inequality is satisfied:
\begin{align}
&L^{\dag}(\{{\bm \beta}_s\}_{s=1}^q, {\bm{U}}, \{{\bm v}_\ell\}_{\ell=1}^k)  \nonumber\\
= & 
\sum_{s=1}^q \sum_{i<j}|e_{is} -e_{js}| 
+\lambda\sum_{s=1}^q\sum_{h=1}^p\left(|\beta_{sh}|-\epsilon\log \left(1+\frac{|\beta_{sh}|}{\epsilon}\right)\right) 
 + \frac{\gamma}{2} \sum_{s=1}^q \sum_{\ell=1}^k u_{s\ell}\|{\bm X}{\bm \beta}_s - {\bm X}{\bm v}_\ell\|_F^2 \nonumber\\
\leq & 
 \sum_{s=1}^q \sum_{o=1}^{n(n-1)/2} w_{os}{|g_{os}-{\bm r}_o^T{\bm \beta}_s|^2}
+\lambda\sum_{s=1}^q\|{\bm \Psi}_s{\bm \beta}_s\|_F^2 +C
+ \frac{\gamma}{2}\sum_{\ell=1}^k \sum_{s=1}^q u_{s\ell}\|{\bm X}{\bm \beta}_s - {\bm X}{\bm v}_\ell\|_F^2 \nonumber\\
 = &  M(\{{\bm \beta}_s\}_{s=1}^q,{\bm U}, \{{\bm v}_\ell\}_{\ell=1}^k|\{{\bm \beta}_{s}^{(t)}\}_{s=1}^q)
\label{mm}
\end{align}
where the equality holds if ${\bm \beta}_{s} = {\bm \beta}_{s}^{(t)}$ in all $(s,h)\; (s=1,2, \cdots, q; \ h =1, 2, \cdots, p)$. 
\end{prop}

From Eq. (\ref{mm}), the updated formula of the proposed method can be calculated based on the least squared criteria. Therefore, it can be expressed simply.

\subsection{Algorithm and updated formula}

This subsection explains the algorithm and updated formula of the proposed method. 
The proposed method updates ${\bm \beta}_s$, ${\bm U}$, and ${\bm v}_\ell$ based on the alternate least squares criterion \citep{alsc} by using the majorizing function of the proposed method.

In Proposition \ref{prop2}, we show the updated formula of ${\bm \beta}_s$ based on the majorizing function. We then show the procedure for the estimation of ${\bm U}$ and ${\bm v}_\ell$ based on $k$-means \citep{forgy1965}. Subsequently, we present the algorithm of the proposed method. 
%
\begin{prop} \label{prop2}
Given $ {\bm X}, w_{os}, g_{os}, {\bm r}_{o} \; (o=1,2,\cdots, n(n-1)/2; s=1, 2, \cdots, q),$ $ \lambda \; (\lambda>0), \gamma \; (\gamma>0), {\bm\Psi}_s,$ and $u_{s\ell} \; (s=1,2,\cdots, q; \ell =1, 2, \cdots, k),$ the updated formula of ${\bm \beta}_d \  (d=1, 2, \cdots, q)$ is derived as follows:
{\small
\begin{align}
{\bm \beta}_d^{(t+1)} \leftarrow& \left( 
2\left({\sum_{o=1}^{n(n-1)/2}w_{os}^{(t+1)}{\bm r}_o{\bm r}_o^T} \right) 
+2 \left(\lambda{\bm\Psi}_d^{(t+1)T} {\bm \Psi}_d^{(t+1)} \right)
+ \gamma \sum_{\ell =1}^k u_{d\ell}^{(t^\dagger)}
{\bm X}^T{\bm X}
-\gamma \sum_{\ell=1}^k \frac{1}{q_\ell}u_{d\ell}^{(t^\dagger)}
{\bm X}^T{\bm X} \right)^{-1}\nonumber \\
& \left(2
\left(\sum_{o=1}^{n(n-1)/2}w_{os}^{(t+1)}g_{os}{\bm r}_o\right)
+ \gamma \sum_{\ell=1}^k \frac{1}{q_\ell} u_{d\ell}^{(t^\dagger)}
\sum_{m \neq d}^q u_{m\ell}^{(t^\dagger)}
{\bm X}^T{\bm X}{\bm \beta}_m^{(f(m,s))}
 \right)
\end{align}}
where $q_\ell$ indicates the number of ${\bm X}{\bm \beta}_s$ belonging to cluster $\ell$.
${\bm{\beta}}_s$ is assumed to be updated in the order $s= 1, 2, \cdots, d, \cdots, q$ as {\rm {Algorithm}} \ref{alg1}.
${\bm{\beta}}_s^{(t)}$ and ${\bm{\beta}}_s^{(t+1)}$ are coefficients at the $t$th and $(t+1)$th steps, respectively.
The function $f(m,s)$ is defined as $t+1$ when $m < s $, whereas it is defined as $t$ when $m>s$.\\
\end{prop}

Next, we explain the updated rule of ${\bm U}$ and the updated formula of ${\bm v}_\ell \ (\ell = 1, 2, \cdots, k)$. ${\bm U}$ and ${\bm v}_\ell$ are updated in the same manner as the $k$-means.\\
{\bf Update ${\bm{U}}$}

${\bm U}$ is updated for each $s$. 
$u_{s\ell^\dag}$ is updated as
\begin{align}
u_{s\ell^\dagger}^{(t^{\dagger}+1)} \leftarrow 
\begin{cases}
1 & \big(\|{\bm X}{\bm \beta}_s^{(t)} - {\bm X}{\bm v}_{\ell^\dag}^{(t^{\dagger})}\|_F^2 \leq 
\|{\bm X}{\bm \beta}_s^{(t)} - {\bm X}{\bm v}_\ell^{(t^{\dagger})}\|_F^2 \ 
\rm{for} \ \rm{any} \ \ell \ (\ell=1, 2, \cdots, k)\big)\\
0 & ({\rm{otherwise}})
\end{cases}
\label{updu}
\end{align}
which is applied to all $s \; (s=1,2,\cdots, q)$ and $\ell^{\dagger} \ (\ell^{\dagger} = 1, 2, \cdots, k$). \\

{\bf Update} ${\bm v}_\ell \ (\ell = 1, 2, \cdots, k)$

${\bm v}_\ell$ is updated as
\begin{align}
{\bm v}_\ell^{(t^{\dagger}+1)} 
\leftarrow \frac{1}{q_\ell}\sum_{s=1}^q u_{s\ell}^{(t^{\dagger}+1)}{\bm \beta}_s^{(t)} \quad
(\ell = 1, 2, \cdots, k).
\label{updvl}
\end{align}
Our proposed method updates ${\bm \beta}_s$, auxiliary variables ${\bm w}_{os}$ and ${\bm \Psi}_s$, indicator matrix ${\bm U}$, and cluster centroid coefficient $\bm{v}_\ell$ alternatively based on the MM algorithm. 
The parameter $k$ is determined by cross validation.
The details of the MM algorithm are described in Algorithm 1.

{\footnotesize
\begin{spacing}{0.01}
\begin{algorithm}[H]
    \caption{Wilcoxon-type MCEN based on MM algorithm}
    \algsetup{linenosize=\small}
\footnotesize
    \label{alg1}
    \begin{algorithmic}[1]    
    \REQUIRE ${\bm X}, {\bm Y}, k ,\; \lambda > 0, \gamma > 0, \;\epsilon > 0;$ threshold for this algorithm $\epsilon^* > 0$
    \ENSURE ${\bm \beta}_s (s=1,2, \cdots, q), {\bm U}, {\bm v}_\ell (\ell = 1, 2, \cdots, k)$\\
    \STATE Set $t \leftarrow 1$ and $t^{\dag} \leftarrow 1$
    \FOR {$s=1$ to $q$}
    \STATE Set initial values ${\bm \beta_s^{(t)}}$
    \ENDFOR \\
    \STATE Set initial ${\bm U}^{(t^{\dag})}$ and ${\bm v}_\ell^{(t^{\dag})} (\ell=1,2,\cdots, k)$ by $k$-means
    \WHILE{ $L^{\dag}(\{{\bm \beta}_s^{(t)}\}_{s=1}^q, {\bm{U}}^{(t^{\dag})}, \{{\bm v}_\ell^{(t^{\dag})}\}_{\ell=1}^k) - L^{\dag}(\{{\bm \beta}_s^{(t+1)}\}_{s=1}^q, {\bm{U}}^{(t^{\dag}+1)}, \{{\bm v}_\ell^{(t^{\dag}+1)}\}_{\ell=1}^k) \geqq \epsilon^*  $}
    \WHILE{ $L^{\dag}(\{{\bm \beta}_s^{(t)}\}_{s=1}^q, {\bm{U}}^{(t^{\dag})}, \{{\bm v}_\ell^{(t^{\dag})}\}_{\ell=1}^k) - L^{\dag}(\{{\bm \beta}_s^{(t+1)}\}_{s=1}^q, {\bm{U}}^{(t^{\dag})}, \{{\bm v}_\ell^{(t^{\dag})}\}_{\ell=1}^k) \geqq \epsilon^*  $}
    \FOR {$s=1$ to $q$}
    \FOR {$o = 1$ to $(n(n-1))/2$}
    \STATE $w_{os}^{(t+1)} \leftarrow 1/ (2|g_{os}-{\bm r}_o^T{\bm \beta}_s^{(t)}|)$
    \ENDFOR \\
    \STATE ${\bm \Psi}_s^{(t+1)} \leftarrow  {\rm diag} \left( \frac{1}{\sqrt{2(\beta_{s1}^{(t)}+\epsilon)} }, 
    \frac{1}{\sqrt{2(\beta_{s2}^{(t)}+\epsilon)} },
    \cdots, \frac{1}{\sqrt{2(\beta_{sp}^{(t)}+\epsilon)} }
    \right)$
    \STATE ${\bm \beta}_s^{(t+1)} \leftarrow \Bigl(
    2\left({\sum_{o=1}^{n(n-1)/2}w_{os}^{(t+1)}{\bm r}_o{\bm r}_o^T} \right) 
    +2 \left(\lambda{\bm\Psi}_s^{(t+1)T} {\bm \Psi}_s^{(t+1)}  \right)$\\
     \ \ \ \ \ \ \ \ \ \ \  \ \ \ \  \ \ \ $+ \gamma \sum_{\ell =1}^k u_{s\ell}^{(t^{\dag})}{\bm X}^T{\bm X}
    -\gamma \sum_{\ell=1}^k \frac{1}{q_\ell}u_{s\ell}^{(t^{\dag})}{\bm X}^T{\bm X}\Bigl)^{-1}$\\
    \ \ \ \ \ \ \ \ \ \ \  \ \ \ \  \ \ \ \ $\times \left(2
    \left(\sum_{o=1}^{n(n-1)/2}w_{os}^{(t+1)}g_{os}{\bm r}_o\right)
    + \gamma \sum_{\ell=1}^k \frac{1}{q_\ell} u_{s\ell}^{(t^{\dag})} \sum_{m \neq s}^q u_{m\ell}^{(t^{\dag})}{\bm X}^T{\bm X}{\bm \beta}_m^{(f(m,s))}
    \right)$
    \ENDFOR \\
    \STATE $t \leftarrow t+1 $
    \ENDWHILE \\
    \WHILE{ $L^{\dag}(\{{\bm \beta}_s^{(t)}\}_{s=1}^q, {\bm{U}}^{(t^{\dag})}, \{{\bm v}_\ell^{(t^{\dag})}\}_{\ell=1}^k) - L^{\dag}(\{{\bm \beta}_s^{(t)}\}_{s=1}^q, {\bm{U}}^{(t^{\dag}+1)}, \{{\bm v}_\ell^{(t^{\dag}+1)}\}_{\ell=1}^k) \geqq \epsilon^*$}
    \FOR{$s=1$ to $q$}
    \STATE Update ${\bm U}^{(t^{\dag})}$ to ${\bm U}^{(t^{\dag}+1)}$ based on Eq. (\ref{updu})
    \ENDFOR
    \FOR{$\ell=1$ to $k$}
    \STATE Update ${\bm v}_\ell^{(t^{\dag})}$ to ${\bm v}_\ell^{(t^{\dag}+1)}$ based on Eq. (\ref{updvl})
    \ENDFOR
    \STATE $t^{\dag} \leftarrow t^{\dag}+1 $
    \ENDWHILE
    \ENDWHILE\\
    \end{algorithmic}
\end{algorithm}
\label{alg}
\end{spacing}
}

\section{Numerical Simulation}

In this section, we demonstrate the efficiency of the proposed method through three numerical simulations. 

\subsection{Simulation design of simulation 1}

This simulation generates covariate matrix ${\bm X}$ and response variables ${\bm Y}$ with a true coefficient matrix. 
The simulation evaluates the prediction accuracy and mean squared error (MSE) of the coefficient matrix.
In this simulation, we use {\choosefont{pcr} RStudio Version 1.4.1103}.

We generate the simulation data, adjusted based on \cite{MCEN}.
The sample size is $50$, and the number of response variables is $9$.
We set the number of covariate matrix $p$ as $12$ and $100$, which will be discussed in Factor 1.2. 
First, let 
\begin{align}
\tilde{\bm \Sigma} = (\tilde{\sigma}_{j'j}), \tilde{\sigma}_{jj}=1, \tilde{\sigma}_{j'j}=0.7, (j',j=1,2,\cdots, 12). 
\label{sig1}
\end{align}
The covariate vectors are generated by ${\bm X}_i \sim \mathcal{N}({\bm 0}_{12}, \tilde{\bm \Sigma})$ for $p=12$.
In the case of $p=100$, ${\bm X}_i \sim \mathcal{N}({\bm 0}_{100}, {\bm \Sigma}^\dag)$, where
\begin{align}
{\bm \Sigma}^\dag 
= \left(
\begin{array}{ll}
\tilde{\bm \Sigma} & {\bm O_{12,88}}\\
{\bm O_{88,12}} & {\bm I}_{88}
\end{array}
\right).
\label{sig2}
\end{align}
${\bm O_{12,88}}$ is a zero matrix with $12$ rows and $88$ columns, ${\bm O_{88,12}}$ is a zero matrix with $88$ rows and $12$ columns, and ${\bm I}_{88}$ is the identity matrix.

To make the matrix of ${\bm B}$, in the case of $p=12$,
we set the ${\bm b}_4 (\eta, \xi) = ({\bm \eta}_4-\xi, {\bm \eta}_4, {\bm \eta}_4+\xi)\in \mathbb{R}^{4\times 3}$.
${\bm \eta}_K$ is a vector with the length of $K$, whose elements are all $\eta$. The determination of the values of $\eta$ and $\xi$ is explained in Factor 1.3 and Factor 1.4, respectively.

Here, the true ${\bm B}^*_{\eta,\xi}$ in $p=12$ is set as follows:
\begin{align}
{\bm B}^*_{\eta, \xi}
=
\left(
\begin{array}{lll}
{\bm b}_4(\eta, \xi) & {\bm O}_{4,3} & {\bm O}_{4,3}\\
{\bm O}_{4,3} & {\bm b}_4(\eta, \xi)  & {\bm O}_{4,3}\\
{\bm O}_{4,3} & {\bm O}_{4,3} & {\bm b}_4(\eta, \xi)
\end{array}
\right).
\label{trueb12}
\end{align}
For the coefficient groups in $p=100$, as is the case of $p=12$, ${\bm b}_{10} (\eta, \xi) = ({\bm \eta}_{10}-\xi, {\bm \eta}_{10}, {\bm \eta}_{10}+\xi)\in \mathbb{R}^{10\times 3}$.
\begin{align}
{\bm B}^*_{\eta, \xi}
=
\left(
\begin{array}{lll}
{\bm b}_{10}(\eta, \xi) & {\bm O}_{10,3} & {\bm O}_{10,3}\\
{\bm O}_{10,3} & {\bm b}_{10}(\eta, \xi)  & {\bm O}_{10,3}\\
{\bm O}_{10,3} & {\bm O}_{10,3} & {\bm b}_{10}(\eta, \xi)\\
{\bm O}_{70,3} & {\bm O}_{70,3} & {\bm O}_{70, 3}\\
\end{array}
\right).
\label{trueb100}
\end{align}
Next, we explain how to generate the response variable, with ${\bm X}$, ${\bm B}$.
The response variable is generated as follows:
\begin{align}
{\bm y}_i = {{\bm B}^*_{\eta, \xi}}^T {\bm X}_i + {\bm  \varepsilon}_i.
\label{ygene}
\end{align}
Here, ${\bm  \varepsilon}_i$ is the random error. 
Several different settings are employed, which are described in detail in Factor 1.5. 
The number of total patterns in this simulation setting is $4$ (Factor 1.1) $\times 2$ (Factor 1.2) $\times 4$ (Factor 1.3) $\times 3$ (Factor 1.4) $\times 4$ (Factor 1.5) $= 384$. For each pattern combination, we generate $50$ learning samples and $1000$ test samples from Eq. (\ref{ygene}), and we repeat the calculation $100$ times.

Next, we explain the evaluation index. In this simulation, we evaluate the prediction variable and coefficient matrix. The median of the absolute prediction error (APE) is
${\rm median} \big \{ |y^*_{is}-\hat{y}_{is}|, \ i=1,2,\cdots,1000; s=1,2,\cdots, 9\big \}$
where $y_{is}^*$ and $\hat{y}_{is}$ represent the testing samples and the prediction variables, respectively.
We also compare the mean squared error of estimator
$(1/(p \times 9)) \sum_{s=1}^9 \| {\hat{\bm \beta}}_s - {\bm \beta}_s^* \|^2_2$,
where ${\bm \beta}_s^*$ is the true ${\bm \beta}_s$, and ${\hat{\bm \beta}_s}$ are the predicted coefficient vectors.
In addition to them, the bias of $\bm{\beta}_s$ are evaluated to assess the estimation accuracy against the compared methods, which can be calculated as bias $=(1/(p \times 9)) \sum_{s=1}^9 \sum_{j^\dagger=1}^p ({\hat{\beta}}_{sj^\dagger} - {\beta}_{sj^\dagger}^*)$, where ${\bm{\beta}}_s = (\beta_{sj^\dagger}^*)$ and $\hat{\bm \beta}_s = (\hat{\beta}_{sj^\dagger}) (j^{\dagger}=1, 2, \cdots, p)$ are the true ${\beta}_{sj^\dagger}$ and predicted ${\beta}_{sj^\dagger}$, respectively.
Spearman's correlation coefficients between $vec(\hat{\bm B}_{\eta, \xi}))$ and $vec ({\bm B}^*_{\eta, \xi})$ are also evaluated to ensure the performance of estimating $\bm{\beta}_s$, where  $\hat{\bm B}_{\eta, \xi}$ represents estimated coefficient matrix, ${\bm B}^*_{\eta, \xi}$ represents true coefficient matrix, and $vec()$ represents vec function. 
\\

\noindent
{\bf Factor 1.1: Method}

We apply the $4$ methods to compare their performance.
In addition to the proposed method, we apply three methods for the comparison: the multivariate cluster elastic net (MCEN) \citep{MCEN}, separate elastic net \citep{elastic}, and Wilcoxon-type lasso \citep{wlasso}. 
The optimization problem of MCEN is shown as follows:
\begin{align*}
\frac{1}{2n}\|{\bm Y}-{\bm X}{\bm B}\|_F^2
+\lambda \sum^{q}_{s=1}\|{\bm \beta}_s\|_1
+ \frac{\gamma}{2} \sum_{s=1}^q \sum_{\ell=1}^k u_{s\ell}\|{\bm X}{\bm \beta}_s - {\bm X}{\bm v}_\ell\|_2^2
\end{align*}
where $k$ is the number of clusters, and $\lambda \ (\lambda >0)$ and $\gamma \ (\gamma >0)$ are the tuning parameters. 
${\bm U} = (u_{s\ell}) \ (s=1,2,\cdots, q; \ell = 1, 2, \cdots, k)$ is the indicator matrix representing the degree of belonging to cluster $\ell$ on the response variable $s$. 
$\bm{v}_\ell \; (\ell = 1, 2, \cdots, k$) is the partial regression coefficient for the cluster centroid.
Elastic net is applied to each response variable separately, denoted as SEN. The objective function of elastic net is as follows: 
\begin{align*}
\| {\bm y}_s - {\bm X}{\bm \beta}_s\|^2_2 + \delta_s\|{\bm \beta}_s\|_1+\gamma_s\|{\bm\beta}_s\|^2_2 \quad (s=1,2,\cdots,q)
\end{align*}
where $\delta_s \; (\delta_s >0)$ and $\gamma_s \; (\gamma_s>0)$ are tuning parameters. 
The third compared method is lasso regression, for which the loss function is a Wilcoxon-type regression function. It is denoted by WLASSO. The objective function of WLASSO is as follows:
\begin{align*}
\sum_{i<j}|e_{is}-e_{js}| + \gamma_s\|{\bm\beta}_s\|_1 \quad (s=1,2,\cdots,q)
\end{align*}
where $e_{is}=y_{is}-\bm{x}_i^T\bm{\beta}_s, e_{js}=y_{js}-\bm{x}_j^T\bm{\beta}_s$, and $\gamma_s \; (\gamma_s>0)$ is the tuning parameter.
MCEN uses the {\choosefont{pcr}mcen} package in R software \citep{MCENr}, and SEN are fitted by the {\choosefont{pcr}glmnet} package in R software \citep{glmnet}. For WLASSO, we use {\choosefont{pcr}LADlasso} \citep{wang2007robust} in the package called {\choosefont{pcr}MTE} \citep{MTE}, as mentioned in \cite{wlasso}.
The tuning parameters of all methods are chosen by five-fold cross validation, and the criteria for cross validation are determined by the minimum median of APE.
For the number of cluster $k$, $2$ and $3$ are candidates for the proposed method and MCEN. \

\noindent
{\bf Factor 1.2: Covariate variable}

The number of the covariate vectors is set as $p= 12$ and $100$. 
The covariate matrix for $p=12$ is described in Eq. (\ref{sig1}), and it is described for $p=100$ in Eq. (\ref{sig2}).

\noindent
{\bf Factor 1.3: Parameter $\eta$ for true coefficient matrix ${\bm B}^*_{\eta, \xi}$}

$\eta$ is a factor to control each value of the non-zero part of a true coefficient matrix. 
The candidates for $\eta$, one of the parameters for the true coefficient matrix, are set as $0.25, 0.5, 0.75$, and $1$.

\noindent
{\bf Factor 1.4: Parameter $\xi$ for true coefficient matrix ${\bm B}^*_{\eta, \xi}$}

$\xi$ is a factor to control the difference between each non-zero true coefficient vector in ${\bm B}^*_{\eta, \xi}$.
The candidates for $\xi$, another parameter for the true coefficient matrix, are set as $0.02, 0.05$, and $0.10$.
With $\eta$ and $\xi$, the true coefficient matrix generates ${\bm B}^*_{\eta, \xi}$ in Eq. (\ref{trueb12}) and Eq. (\ref{trueb100}).

\noindent
{\bf Factor 1.5: Error distribution}

We set four different patterns for ${\bm \varepsilon}_i = (\varepsilon_{is})$ based on \cite{wlasso}.
Error 1 is set as 
$\varepsilon_{is} \sim N(0, 1) \ (i=1,2,\cdots,n; s=1,2,\cdots, q).$ 
Error 2 is set as 
$\varepsilon_{is} \sim 0.95N(0,1)+0.05N(0,100)  \ (i=1,2,\cdots,n; s=1,2,\cdots, q).$
Error 3 is set as 
$\varepsilon_{is} \sim \sqrt{2}t(4)  \ (i=1,2,\cdots,n; s=1,2,\cdots, q)$, 
where $t(4)$ represents the $t$ distribution with $4$ degree of freedom. 
Error 4 is set as 
$\varepsilon_{is} \sim Cauchy(0,1)  \ (i=1,2,\cdots,n; s=1,2,\cdots, q)$, 
where $Cauchy(0,1)$ represents a Cauchy distribution with a location parameter of $0$ and a scale parameter of $1$.
Error 1 assumes a normal distribution, Error 2 assumes a distribution containing outliers, and Error 3 and Error 4 assume heavy-tailed distribution.\\

\subsection{Simulation design of simulation 2}

We also conduct simulations to examine how the estimated values change with the magnitude of the noise in error distribution. In this simulation, the covariate vectors are generated by ${\bm X}_i \sim \mathcal{N}({\bm 0}_{50}, {\bm \Sigma}^\dagger $) for $p=50$, where
\begin{align}
{\bm \Sigma}^\dagger
= \left(
\begin{array}{ll}
\tilde{\bm \Sigma} & {\bm O_{12,38}}\\
{\bm O_{38,12}} & {\bm I}_{38}
\end{array}
\right).
\label{sig3}
\end{align}
The matrix of $\bm{B}$, ${\bm b}_{10} (\eta, \xi) = ({\bm \eta}_{10}-\xi, {\bm \eta}_{10}, {\bm \eta}_{10}+\xi)\in \mathbb{R}^{10\times 3}$, which is same as $p=100$, and the true ${\bm B}^*_{\eta, \xi}$ for $p=50$ is
\begin{align}
{\bm B}^*_{\eta, \xi}
=
\left(
\begin{array}{lll}
{\bm b}_{10}(\eta, \xi) & {\bm O}_{10,3} & {\bm O}_{10,3}\\
{\bm O}_{10,3} & {\bm b}_{10}(\eta, \xi)  & {\bm O}_{10,3}\\
{\bm O}_{10,3} & {\bm O}_{10,3} & {\bm b}_{10}(\eta, \xi)\\
{\bm O}_{20,3} & {\bm O}_{20,3} & {\bm O}_{20, 3}\\
\end{array}
\right),
\label{trueb50}
\end{align}
where $\xi = 0.5$. The generation of the response variables is same as Eq. (\ref{ygene}). The setting of ${\bm \varepsilon}_i$ is explained in Factor 2.3. The number of total patterns in this simulation is $4$ (Factor 2.1) $\times 4$ (Factor 2.2) $\times 5$ (Factor 2.3) $= 80$. For each pattern combination, we generate $100$ learning samples and $100$ test samples from Eq. (\ref{ygene}), and we repeat the calculation $100$ times, which is the same as the first simulation setting.\\
For the evaluation index of this simulation, we employ the mean of median APE.

Next, we explain each Factor for the second simulation.\\

\noindent
{\bf Factor 2.1: Method}

The compared methods are the  same as those in the first simulation: MCEN, SEN, and WLASSO.

\noindent
{\bf Factor 2.2: Parameter $\eta$ for true coefficient matrix ${\bm B}^*_{\eta, \xi}$}

The candidates for $\eta$, one of the parameters for the true coefficient matrix, are set as $0.25, 0.5, 0.75$, and $1$, which are the same as those used in the first simulation.

\noindent
{\bf Factor 2.3: Proportion of contaminated distribution}

We set five different patterns of noise rate for ${\bm \varepsilon}_i = (\varepsilon_{is}) \ (i=1,2,\cdots,n; s=1,2,\cdots, q)$. The proportion of noise increases from Error 1$^\dagger$ to Error 5$^\dagger$.\\
Error 1$^\dagger$ : 
$\varepsilon_{is} \sim N(0, 1)$, \\
Error 2$^\dagger$ :
$\varepsilon_{is} \sim 0.95N(0,1)+0.05N(0,100)$,\\
Error 3$^\dagger$  
$\varepsilon_{is} \sim 0.9N(0,1)+0.1N(0,100)$,\\ 
Error 4$^\dagger$ : $\varepsilon_{is} \sim 0.85N(0,1)+0.15N(0,100)$, and\\
Error 5$^\dagger$ : $\varepsilon_{is} \sim 0.8N(0,1)+0.2N(0,100)$.\\

\subsection{Simulation design of simulation 3}
In this subsection, we explain another simulation to verify the relationship between sample size and MSE of $\bm{\beta}_s$. The simulation is performed to examine whether the value of the MSE of $\bm{\beta}_s$ tends to decrease as the sample size increases. We set sample size as $25, 50, 75$ and $100$, and the number of the covariate matrix $p$ as $12$ and $50$. The covariate vectors are generated same as those in simulations $1$ and $2$. The true $\bm{B}_{\eta, \xi}^*$ for $p=12$ and $p=50$ are set in the same way of Eq. (\ref{trueb12}) and Eq. (\ref{trueb50}), respectively. In this simulation, we set $\xi = 0.05$. The response variable is generated by Eq. (\ref{ygene}), where the random error $\bm{\varepsilon}_i \sim N(0, 1)$.
The number of total patterns in this simulation setting is $2$ (Factor 3.1) $\times 3$ (Factor 3.2)  $= 6$. For each pattern combination, we generate $100$ learning samples from Eq. (\ref{ygene}), and repeat calculating the MSE of $\bm{\beta}_s$ $100$ times.

For the evaluation index, we adapt the mean of the MSE of $\bm{\beta}_s$.\\

\noindent
{\bf Factor 3.1: Covariate variable}

The covariate matrix for $p=12$ is described in Eq. (\ref{sig1}), and it is described for $p=50$ in Eq. (\ref{sig3}).

\noindent
{\bf Factor 3.2: Parameter $\eta$ for true coefficient matrix ${\bm B}^*_{\eta, \xi}$}

The candidates for $\eta$, one of the parameters for the true coefficient matrix, are set as $0.5, 0.75$, and $1$.\\

\subsection{Simulation result}

The results of the simulation $1$ are displayed by error distribution in Table \ref{p12er1} to Table \ref{bico4}. Table \ref{p12er1} to Table \ref{p12er4} show the results of median APE and MSE of $\bm{\beta}_s$, and Table \ref{bico1} to Table \ref{bico4} demonstrate those of bias of $\bm{\beta}_s$ and Spearman's correlation coefficients of $\bm{B}$. We also present the plots of all these results in ${\rm Appendix \ D}$ in Supplementary material. We start by considering the results of the median APE and MSE of $\bm{\beta}_s$. 
First, as observed in each outcome for Factor 1.1, {the results of both the median APE and the MSE of $\bm{\beta}_s$ of the proposed method were better than those of the other compared methods in almost all patterns of ($\eta$, $\xi$) in Error 2 (Table \ref{p12er2}) and Error 4 for both $p=12$ and $100$ (Table \ref{p12er4}). 
Focusing on the median APE, the proposed method was better in all patterns of ($\eta$, $\xi$) in Error 3 for $p=12$ (Table \ref{p12er3}); Error 1 for $p=12$ (Table \ref{p12er1}), except $\eta = 0.25$; and Error 2 of $p=12$ (Table \ref{p12er2}), except $(\eta, \xi) = (0.75, 0.05)$. 
In terms of the MSE of ${\bm \beta}_s$ (the second evaluation index), the proposed method had the smallest values in all patterns except $(\eta, \xi) = (0.75, 0.05)$ in Error 2 for $p=12$ (Table \ref{p12er2}); and Error 3 for $p=12$ except $\eta = 0.25$. 
For the compared methods, the MCEN had smaller values in almost all patterns of Error 3 for  $p=100$ in both evaluation indices. 
For the median APE, the MCEN was better in several patterns of Error 1 for $p=100$ (Table \ref{p12er1}) as well as 
in Error 1 for $p=12$ and $100$ for the MSE of ${\bm \beta}_s$. 
SEN was better for several patterns of $\eta = 0.25$ of Error 1 for $p=12$ (Table \ref{p12er1}) for both evaluations indices as well as in Error 1 for $p=100$ (}Table \ref{p12er1}) for the median APE. 
WLASSO was only better than the other methods in $\eta=0.75$ and $\xi=0.05$ in Error 2 for $p=12$ (Table \ref{p12er2}), however, compared with the proposed method, the difference in the values was small in Error 2.

\begin{table}
\begin{center}
\caption{Result of the median APE and MSE of $\bm{\beta}_s$ in $p=12$ and $p=100$ for Error $1$}\label{p12er1}
\scalebox{0.7}{
\begin{tabular}{c c r rrr rrr} 
\hline
\toprule
& \quad & &  \multicolumn{3}{c}{the median of APE (sd of APE)} &  \multicolumn{3}{c}{MSE of ${\bm \beta}_s$ (sd of MSE)}  \\ \cmidrule(lr){4-6} \cmidrule(lr){7-9}
& & $\xi$ & 0.02 & 0.05 & 0.10 & 0.02 & 0.05 & 0.10\\
$p$ & $\eta$ & & & & & & &\\ 
\midrule
12 & 0.25 & WMCEN & 0.708 (0.010) & 0.708 (0.010) & $\bm{0.705 (0.010)}$ & 0.023 (0.004) & 0.023 (0.004) &  0.023 (0.004)\\
\multirow{11}{*}{ } & & MCEN & 0.712 (0.011) &  0.712 (0.011) & 0.715 (0.014) & 0.018 (0.005)  & $\bm{0.017 (0.004) }$ & 0.018 (0.003)\\
&   & SEN & $\bm{0.705}$ $\bm{(0.010)}$ & $\bm{0.707 (0.011)}$ & $0.706 (0.011)$ & $\bm{0.018 (0.003)}$ & 0.019 (0.003) & $\bm{0.018 (0.004)}$\\
&   & WLASSO & 0.724 (0.011) & 0.723 (0.011) & 0.717 (0.010) & 0.036 (0.007) & 0.035 (0.007) & 0.030 (0.006)\\
&  0.50 & WMCEN & $\bm{0.722 (0.010)}$ & $\bm{0.725 (0.011)}$ & $\bm{0.723 (0.010)}$ & 0.034 (0.006) & 0.037 (0.006) & 0.034 (0.006)\\
&  & MCEN & 0.735 (0.022) & 0.735 (0.023) & 0.741 (0.023) & $\bm{0.028 (0.007)}$ & $\bm{0.028 (0.007)}$  & $\bm{0.031}$ $\bm{(0.007)}$\\
&  & SEN & 0.726 (0.012) & 0.726 $\bm{(0.012)}$ & 0.726 (0.012) & 0.035 (0.006) & 0.035 (0.006) & 0.034 (0.006)\\
&  & WLASSO & 0.732 (0.011) & 0.736 (0.012) & 0.736 (0.012) & 0.043 (0.007) & 0.047 (0.008) & 0.046 (0.008)\\
&  0.75 & WMCEN & $\bm{0.725 (0.011)}$ & $\bm{0.725 (0.011)}$  & $\bm{0.727 (0.011)}$ & 0.038 (0.007) & 0.038 (0.008) & 0.039 (0.008)\\
&  & MCEN & 0.752 (0.043) & 0.753 (0.044) & 0.754 (0.044) & $\bm{0.027 (0.007)}$ & $\bm{0.027 (0.007)}$ & $\bm{0.027 (0.007)}$\\
&  & SEN & 0.738 (0.012) & 0.733 (0.012) & 0.731 (0.013) & 0.046 (0.008) & 0.042 (0.007) & 0.039 (0.007)\\
&  & WLASSO & 0.740 (0.013) & 0.740 (0.013) & 0.740 (0.013) & 0.051 (0.009) & 0.051 (0.009) & 0.051 (0.009)\\
&  1.00 & WMCEN & $\bm{0.727 (0.012)}$ & $\bm{0.727 (0.012)}$ & $\bm{0.727 (0.011)}$ & 0.039 (0.008) & 0.039 (0.008) & 0.039 (0.008)\\
&  & MCEN & 0.783 (0.073) & 0.781 (0.071) & 0.781 (0.071) & $\bm{0.032 (0.010)}$ & $\bm{0.032 (0.009)}$  & $\bm{0.032 (0.009)}$ \\
&  & SEN & 0.736 (0.011) & 
0.739 (0.012) & 0.737 (0.014) & 0.047 (0.008) & 0.044 (0.007) & 0.048 (0.008)\\
&  & WLASSO & 0.741 (0.013) & 0.741 (0.013) & 0.741 (0.013) & 0.052 (0.010) & 0.052 (0.010) & 0.052 (0.010)\\
 \hline
 &   & $\xi$ & 0.02 & 0.05 & 0.10 & 0.02 & 0.05 & 0.10 \\
$p$ & $\eta$ &  &  &  &  &  &  &  \\\hline
100 & 0.25 & WMCEN & 0.848 (0.016) & 0.844 (0.017) & 0.844 (0.017) & 0.007 (<0.001) & 0.007 (<0.001) & 0.007 (<0.001) \\
 &  & MCEN & $\bm{0.846 (0.027)}$ & 0.848 (0.027) & $\bm{0.808 (0.023)}$ & $\bm{0.005 (< 0.001)}$ & $\bm{0.005 (< 0.001)}$ & $\bm{0.005 (< 0.001)}$ \\
 &  & SEN & $\bm{0.846 (0.016)}$ & $\bm{0.817 (0.014)}$ & 0.826 (0.015) & 0.007 (<0.001) & $\bm{0.005 (< 0.001)}$  & 0.006 (<0.001) \\
 &  & WLASSO & 1.111 (0.034) & 1.111 (0.035) & 1.100 (0.035) & 0.020 (0.002) & 0.020 (0.002) & 0.020 (0.002) \\
 & 0.50 & WMCEN & 0.968 (0.033) & 0.999 (0.032) & 1.002 (0.33) & 0.014 (0.002) & 0.015 (0.002) & 0.015 (0.002) \\
 &  & MCEN & $\bm{0.926 (0.054)}$ & $\bm{0.920 (0.054)}$ & $\bm{0.910 (0.053)}$ & $\bm{0.009 (0.002)}$ & $\bm{0.009 (0.002)}$ & $\bm{0.009 (0.002)}$ \\
 &  & SEN & 0.979 (0.031) & 0.964 (0.028) & 1.004 (0.029) & 0.014 (0.002) & 0.013 (0.002) & 0.015 (0.002) \\
 &  & WLASSO & 1.203 (0.040) & 1.197 (0.041) & 1.195 (0.041) & 0.027 (0.003) & 0.027 (0.003) & 0.027 (0.003) \\
 & 0.75 & WMCEN & 1.078 (0.051) & 1.078 (0.051) & 1.081 (0.050) & 0.021 (0.003) & 0.021 (0.003) & 0.021 (0.003) \\
 &  & MCEN & 1.068 (0.093) & 1.068 (0.092) & $\bm{1.068 (0.093)}$ & $\bm{0.016 (0.003)}$ & $\bm{0.016 (0.003)}$ & $\bm{0.016 (0.003}$ \\
 &  & SEN & $\bm{1.045 (0.050)}$ & $\bm{1.044 (0.044)}$ & 1.070 (0.046) & 0.019 (0.003) & 0.019 (0.003) & 0.021 (0.003) \\
 &  & WLASSO & 1.260 (0.057) & 1.260 (0.056) & 1.253 (0.056) & 0.032 (0.004) & 0.032 (0.004)  & 0.032 (0.004)  \\
 & 1.00 & WMCEN & 1.180 (0.064) & 1.183 (0.064) & 1.115 (0.067) & 0.028 (0.005) & 0.028 (0.005)  & 0.024 (0.005)  \\
 &  & MCEN & 1.165 (0.132) & 1.166 (0.132) & 1.165 (0.131) & $\bm{0.021 (0.004)}$ & $\bm{0.021 (0.004)}$ & $\bm{0.021 (0.004)}$ \\
 &  & SEN & $\bm{1.101 (0.066)}$ & $\bm{1.111 (0.065)}$ & $\bm{1.088 (0.063)}$ & 0.024 (0.005) & 0.026 (0.005) & 0.023 (0.005) \\
 &  & WLASSO & 1.285 (0.068) & 1.272 (0.068) & 1.283 (0.068) &0.035 (0.006) & 0.034 (0.005) & 0.035 (0.006) \\\hline
\end{tabular}
}
\end{center}
$\quad \quad \quad \quad \quad \quad \quad$ 
\end{table}

\begin{table}
\begin{center}
\caption{Result of the median APE and MSE of $\bm{\beta}_s$ in $p=12$ and $p=100$ for Error $2$}\label{p12er2}
\scalebox{0.7}{
\begin{tabular}{cc r rrr rrr} 
\hline
\toprule
& \quad  & &  \multicolumn{3}{c}{the median of APE (sd of APE)} &  \multicolumn{3}{c}{MSE of ${\bm \beta}_s$ (sd of MSE)}  \\\cmidrule(lr){4-6} \cmidrule(lr){7-9}
 & & $\xi$ & 0.02 & 0.05 & 0.10 & 0.02 & 0.05 & 0.10\\
$p$ & $\eta$ & & & & & & &\\ 
\midrule
12 & 0.25 & WMCEN & $\bm{0.755 }$ $(0.014)$ & $\bm{0.755 }$ $(0.014)$ & $\bm{0.755}$ $(0.014)$ & $\bm{0.025}$ $\bm{(0.005)}$ & $\bm{0.025 (0.005)}$ & $\bm{0.025 (0.005)}$\\
&  & MCEN & 2.699 (1.018) & 2.629 (1.018) &  2.670 (1.016) & 0.919 (0.845) & 0.513 (0.488) & 0.694 (0.691)\\
& & SEN & 2.274 (1.150) & 2.291 (1.663) & 2.815 (1.541) & 9.714 (8.113) & 15.024 (11.884) & 13.595 (11.206)\\
& & WLASSO & 0.777 (0.019) & 0.776 (0.019) & 0.776 (0.019) & 0.043 (0.012) & 0.043 (0.012) & 0.043 (0.012)\\ 
& 0.50 & WMCEN & $\bm{0.776 (0.018)}$ & $\bm{0.777 (0.018)}$ & $\bm{0.779}$ $\bm{(0.018)}$ & $\bm{0.042 (0.010)}$ & $\bm{0.043}$ $\bm{(0.010)}$ & $\bm{0.043}$ $\bm{(0.010)}$\\
&  & MCEN & 2.717 (1.034) & 2.728 (1.038) & 2.747 (1.048) & 0.767 (0.641) & 0.815 (0.675) & 0.898 (0.737)\\
&  & SEN & 3.284 (1.660) & 3.554 (1.836) & 3.222 (1.651) & 15.791 (12.251) & 17.447 (13.151) & 15.091 (11.791)\\
&  & WLASSO & 0.805 (0.025) & 0.804 (0.025) & 0.805 (0.025) & 0.068 (0.019) &  0.068 (0.019) & 0.068 (0.019)\\
& 0.75 & WMCEN & $\bm{0.790}$ $ \bm{(0.024)}$ & 0.827 (0.032) & $\bm{0.789 (0.022)}$ & $\bm{0.055 (0.017)}$ & 0.089 (0.027) & $\bm{0.052 (0.014)}$\\
& & MCEN & 2.893 (1.058) & 2.945 (1.104) & 2.957 (1.110) & 1.560 (1.211) & 1.896 (1.440) & 1.951 (1.475)\\
& & SEN & 3.399 (1.566) & 3.513 (1.677) & 3.423 (1.637) & 16.332 (12.469) & 17.001 (13.511) & 15.994 (12.151)\\
& & WLASSO & 0.810 (0.027) & $\bm{0.810 (0.027)}$ & 0.809 (0.027) & 0.072 (0.021) & $\bm{0.072 (0.021)}$ & 0.071 (0.021)\\
& 1.00 & WMCEN & $\bm{0.787 (0.025)}$ & $\bm{0.791}$ $\bm{ (0.026)}$ & $\bm{0.790}$ $\bm{ (0.026)}$  & $\bm{0.053 (0.018)}$ & $\bm{0.057 (0.019)}$ & $\bm{0.054}$ $\bm{(0.019)}$\\
&  & MCEN & 2.987 (1.089) & 2.990 (1.091) & 2.998 (1.094) & 1.853 (1.327) & 1.870 (1.340) & 1.898 (1.356)\\
&  & SEN & 3.573 (1.714) & 3.741 (1.841) & 3.431 (1.603) & 16.569 (12.664) & 18.070 (13.953) & 15.150 (11.303)\\
&  & WLASSO & 0.812 (0.029) & 0.812 (0.029) & 0.812 (0.029) & 0.074 (0.024) & 0.074 (0.024) & 0.074 (0.024)\\
 \hline
& & $\xi$ & 0.02 & 0.05 & 0.10 & 0.02 & 0.05 & 0.10\\
$p$ & $\eta$ &  &  & &  & &  & \\\hline
100 & 0.25 & WMCEN & $\bm{0.871 (0.024)}$ & $\bm{0.891 (0.024)}$ & $\bm{0.874 (0.022)}$ & $\bm{0.005 (<0.001)}$ & $\bm{0.006 (0.001)}$ & $\bm{0.005 (0.001)}$ \\
& & MCEN & 5.893 (2.714) & 6.308 (2.863) & 5.290 (2.382) & 1.141 (0.681) & 1.1357 (0.846) & 0.888 (0.524) \\
& & SEN & 10.536 (6.366) & 11.010 (6.620) & 10.019 (6.162) & 5.240 (4.030) & 5.576 (4.226) & 4.762 (3.708) \\
& & WLASSO & 1.718 (0.940) & 1.716 (0.937) & 1.718 (0.942) & 0.170 (0.437) & 0.169 (0.437) & 0.170 (0.437) \\
 & 0.50 & WMCEN & $\bm{1.087 (0.049)}$ & $\bm{1.090 (0.052)}$ & $\bm{1.088 (0.048)}$ & $\bm{0.016 (0.002)}$ & $\bm{0.016 (0.002)}$ & $\bm{0.016 (0.002)}$ \\
& & MCEN & 7.287 (3.411) & 7.295 (3.416) & 7.301 (3.422) & 1.768 (1.085) & 1.772 (1.087) & 1.792 (1.095)\\
& & SEN & 11.232 (6.546) & 10.761 (6.334) & 11.375 (6.601) & 5.545 (4.160) & 5.190 (3.945) & 5.673 (4.2560) \\
& & WLASSO & 2.028 (1.115) & 2.166 (1.365) & 2.23 (1.463) & 0.270 (0.599) & 0.371 (0.793) & 0.389 (0.814) \\
 & 0.75 & WMCEN & $\bm{1.285 (0.112)}$ & $\bm{1.245 (0.095)}$ & $\bm{1.224 (0.086)}$ & $\bm{0.028 (0.006)}$ & $\bm{0.026 (0.005)}$ & $\bm{0.026 (0.005)}$ \\
&  & MCEN & 9.525 (4.610) & 9.520 (4.596) & 9.528 (4.585) & 3.145 (2.011) & 3.143 (2.011) & 3.143 (2.011) \\
& & SEN & 11.895 (6.747) & 11.519 (6.682) & 11.421 (6.522) & 6.049 (4.485) & 5.802 (4.364) & 5.802 (4.175) \\
& & WLASSO & 2.434 (1.405) & 2.433 (1.405) & 2.398 (1.203) & 0.423 (0.774) & 0.422 (0.773) & 0.422 (0.583) \\
 & 1.00 & WMCEN & $\bm{1.596 (0.323)}$ & $\bm{1.609 (0.300)}$ & $\bm{1.606 (0.305)}$ & $\bm{0.055 (0.029)}$ & $\bm{0.057 (0.028)}$ & $\bm{0.057 (0.028)}$ \\
& & MCEN & 10.07 (4.807) & 10.069 (4.805) & 10.066 (4.809) & 3.443 (2.236) & 3.442 (2.236) & 3.442 (2.237) \\
& & SEN & 11.785 (6.666) & 11.590 (6.533) & 11.962 (6.800) & 5.933 (4.358) & 5.698 (4.193) & 6.154 (4.564) \\
& & WLASSO & 2.595 (1.405) & 2.598(1.414) & 2.600 (1.420) & 0.469 (0.808) & 0.470 (0.814) & 0.472 (0.820) \\\hline
\end{tabular}
}
\end{center}
$\quad \quad \quad \quad \quad \quad \quad$ 
\end{table}

\begin{table}
\begin{center}
\caption{Result of the median APE and MSE of $\bm{\beta}_s$ in $p=12$ and $p=100$ for Error $3$}\label{p12er3}
\scalebox{0.7}{
\begin{tabular}{ccr rrr rrr} 
\\ \hline
\toprule
& \quad &  &  \multicolumn{3}{c}{the median of APE (sd of APE)} &  \multicolumn{3}{c}{MSE of ${\bm \beta}_s$ (sd of MSE)}  \\\cmidrule(lr){4-6} \cmidrule(lr){7-9}
& & $\xi$ & 0.02 & 0.05 & 0.10 & 0.02 & 0.05 & 0.10\\
$p$ & $\eta$ & & & & & & &\\ 
\midrule
12 & 0.25 & WMCEN & $\bm{1.097}$  $\bm{(0.016)}$ & $\bm{1.099 (0.017)}$ & $\bm{1.098 (0.017)}$ & 0.038 (0.007) & 0.040 (0.007) & 0.038 (0.007) \\
& & MCEN & 1.121 (0.023) & 1.118 (0.021) &  1.118 (0.021) & $\bm{0.036 (0.014)}$ & $\bm{0.031}$ $\bm{(0.007)}$ & $\bm{0.030}$ $\bm{(0.007)}$ \\
& & SEN & 1.118 (0.022) & 1.118 (0.022) & 1.121 (0.024) & 0.044 (0.021) & 0.038 (0.022) & 0.055 (0.034)\\
& & WLASSO & 1.117 (0.018) & 1.118 (0.018)  & 1.118 (0.018)  & 0.057 (0.011) & 0.058 (0.011) & 0.057 (0.012)\\
& 0.50 & WMCEN & $\bm{1.126 (0.018)}$ & $\bm{1.125 (0.018)}$ & $\bm{1.124 (0.018)}$ & $\bm{0.067}$ $\bm{(0.011)}$ & $\bm{0.067}$ $\bm{(0.011)}$ & $\bm{0.065 (0.011)}$\\
& & MCEN & 1.163 (0.028) & 1.160 (0.028) & 1.164 (0.029) & 0.069 (0.015) & 0.069 (0.018) & 0.068 (0.014) \\
& & SEN & 1.147 (0.023) & 1.154 (0.024) & 1.144 (0.023)  & 0.079 (0.022) & 0.082 (0.021) & 0.074 (0.019)\\
& & WLASSO & 1.155 (0.023) & 1.155 (0.023) & 1.154 (0.023) & 0.099 (0.020) & 0.099 (0.020) & 0.098 (0.020)\\
& 0.75 & WMCEN & $\bm{1.147}$ $\bm{(0.022)}$ & $\bm{1.147}$ $\bm{(0.022)}$  & $\bm{1.152 (0.022)}$ & $\bm{0.090 (0.017)}$ & $\bm{0.090 (0.017)}$ & $\bm{0.094 (0.018)}$\\
& & MCEN & 1.199 (0.045) & 1.200 (0.045) & 1.201 (0.046) & 0.095 (0.023) & 0.096 (0.023) & 0.096 (0.023)\\
& & SEN & 1.184 (0.029) & 1.174 (0.026) & 1.176 (0.026) & 0.115 (0.030) & 0.111 (0.029) & 0.106 (0.025)\\
& & WLASSO & 1.173 (0.027) & 1.172 (0.027) & 1.172 (0.026) & 0.118 (0.025) & 0.118 (0.025) & 0.117 (0.025) \\
& 1.00 & WMCEN & $\bm{1.155 (0.026)}$ & $\bm{1.159 (0.026)}$ & $\bm{1.159}$ $\bm{(0.026)}$ & $\bm{0.100 (0.024)}$ & $\bm{0.104 (0.023)}$ & $\bm{0.104 (0.023)}$\\
& & MCEN & 1.217 (0.068) & 1.230 (0.070) & 1.230 (0.070) & 0.098 (0.039) & 0.108 (0.034) & 0.108 (0.033)\\
& & SEN & 1.197 (0.030) & 1.196 (0.030) & 1.193 (0.030) & 0.134 (0.034) & 0.136 (0.032) & 0.136 (0.035)\\
& & WLASSO & 1.192 (0.031)  & 1.192 (0.031) & 1.191 (0.030) & 0.144 (0.031) & 0.144 (0.031) & 0.144 (0.031)\\\hline
& & $\xi$ & 0.02 & 0.05 & 0.10 & 0.02 & 0.05 & 0.10\\
$p$ & $\eta$ &  &  &  &  &  & &  \\\hline
100 & 0.25 & WMCEN & 1.293 (0.025) & $\bm{1.205 (0.024)}$ & $\bm{1.240 (0.022)}$ & 0.011 (0.001) & $\bm{0.006 (0.001)}$ & 0.009 (0.001) \\
& & MCEN & $\bm{1.256 (0.033)}$ & 1.247 (0.033) & 1.246 (0.036) & $\bm{0.008 (0.001)}$ & 0.007 (0.001) & $\bm{0.007 (0.001)}$ \\
& & SEN & 1.280 (0.038) & 1.262 (0.036) & 1.271 (0.036) & 0.010 (0.002) & 0.010 (0.003) & 0.010 (0.003) \\
& & WLASSO & 1.956 (0.090) & 1.961 (0.093) & 1.941 (0.086) & 0.064 (0.013) & 0.064 (0.013) & 0.062 (0.012) \\
 & 0.50 & WMCEN & 1.420 (0.036) & 1.438 (0.039) & $\bm{1.405 (0.035)}$ & 0.020 (0.002) & 0.022 (0.002) & 0.019 (0.002) \\
& & MCEN & $\bm{1.382 (0.062)}$ & $\bm{1.396 (0.061)}$ & 1.422 (0.060) & $\bm{0.016 (0.003)}$ & $\bm{0.017 (0.003)}$ & $\bm{0.018 (0.003)}$ \\
& & SEN & 1.485 (0.050) & 1.466 (0.046) & 1.469 (0.060) & 0.024 (0.004) & 0.022 (0.003) & 0.024 (0.005) \\
& & WLASSO & 2.049 (0.087) & 2.074 (0.092) & 2.034 (0.082) & 0.075 (0.013) & 0.077 (0.013) & 0.073 (0.012) \\
& 0.75 & WMCEN & 1.584 (0.053) & 1.588 (0.058) & 1.588 (0.062) & 0.033 (0.003) & 0.034 (0.004) & 0.033 (0.004) \\
& & MCEN & $\bm{1.526 (0.097)}$ & $\bm{1.509 (0.098)}$ & $\bm{1.513 (0.097)}$ & $\bm{0.025 (0.005)}$ & $\bm{0.024 (0.005)}$ & $\bm{0.024 (0.005)}$ \\
& & SEN & 1.631 (0.066) & 1.642 (0.070) & 1.664 (0.065) & 0.037 (0.006) & 0.039 (0.006) & 0.039 (0.005) \\
& & WLASSO & 2.195 (0.098) & 2.176 (0.095) & 2.199 (0.105) & 0.093 (0.015) & 0.091 (0.015) & 0.095 (0.017) \\
 & 1.00 & WMCEN & 1.754 (0.085) & 1.754 (0.086) & 1.752 (0.082) & 0.049 (0.007) & 0.049 (0.007) & 0.049 (0.007) \\
& & MCEN & $\bm{1.638 (0.134)}$ & $\bm{1.638 (0.134)}$ & $\bm{1.656 (0.132)}$ & $\bm{0.033 (0.007)}$ & $\bm{0.033 (0.007)}$ & $\bm{0.034 (0.007)}$ \\
& & SEN & 1.761 (0.083) & 1.744 (0.084) & 1.808 (0.083) & 0.050 (0.007) & 0.049 (0.008) & 0.054 (0.008) \\
& & WLASSO & 2.288 (0.120) & 2.273 (0.115) & 2.288 (0.019) & 0.108 (0.019) & 0.105 (0.017) & 0.108 (0.019) \\\hline
\end{tabular}
}
\end{center}
$\quad \quad \quad \quad \quad \quad \quad$ 
\end{table}

\begin{table}
\begin{center}
\caption{Result of the median APE and MSE of $\bm{\beta}_s$ in $p=12$ and $p=100$ for Error $4$ }\label{p12er4}
\scalebox{0.6}{
\begin{tabular}{ccr rrr rrr} 
\toprule
& \quad  & & \multicolumn{3}{c}{the median of APE (sd of APE)} &  \multicolumn{3}{c}{MSE of ${\bm \beta}_s$ (sd of MSE)}  \\\cmidrule(lr){4-6} \cmidrule(lr){7-9}
& & $\xi$ & 0.02 & 0.05 & 0.10 & 0.02 & 0.05 & 0.10\\
$p$ & $\eta$ & & & & & & &\\ 
\midrule
12 & 0.25 & WMCEN & $\bm{1.134}$ $\bm{(0.033)}$ &  $\bm{1.096 (0.028)}$ & $\bm{1.099 (0.028)}$ & $\bm{0.068 (0.019)}$ & $\bm{0.045 (0.012)}$ & $\bm{0.045 (0.012)}$\\
& & MCEN & 2.281 (0.704) &  2.283 (0.707) & 2.283 (0.688) & 7.815 (54.839) & 6.739 (43.106) & 3.745 (17.255)\\
& & SEN & 2.315 (0.665) & 2.362 (0.748) & 2.142 (0.612) & 2847.525 (23988.616) & 2824.287 (23908.141) & 2793.485 (23757.635)\\
& & WLASSO & 1.198 (0.047) & 1.192 (0.046)  & 1.183 (0.043) & 0.135 (0.056) & 0.130 (0.055) & 0.123 (0.054)\\
&  0.50 & WMCEN & $\bm{1.151 (0.032)}$ & $\bm{1.151}$ $\bm{(0.031)}$ & $\bm{1.152}$ $\bm{(0.031)}$  &  $\bm{0.081 (0.017)}$ & $\bm{0.080 (0.016)}$ & $\bm{0.080 (0.017)}$\\
& & MCEN &  2.393 (0.674) & 2.393 (0.676)  & 2.395 (0.679) & 10.727 (64.328) & 10.282 (59.072) & 9.687 (62.089)\\
& & SEN & 2.484 (0.615) & 2.464 (0.613) & 2.542 (0.728) & 2850.145 (23948.576) & 2829.491 (23929.864) & 2841.953 (23942.039) \\
& & WLASSO & 1.245 (0.051) & 1.246 (0.051)  & 1.250 (0.052) & 0.169 (0.060) & 0.169 (0.060) & 0.172 (0.060)\\
&  0.75 & WMCEN & $\bm{1.212}$ $\bm{(0.041)}$ & $\bm{1.194 (0.038)}$  & $\bm{1.196}$ $\bm{(0.037)}$ & $\bm{0.130 (0.029)}$ & $\bm{0.113 (0.023)}$ & $\bm{0.113 (0.024)}$\\
& & MCEN & 2.497 (0.672) & 2.499 (0.674) & 2.498 (0.675) & 9.470 (60.942) & 9.352 (60.420) & 9.177 (59.789)\\
& & SEN & 2.768 (0.726) & 2.778 (0.746) & 2.697 (0.698) & 2856.989 (23963.969) & 2860.519 (23954.087) & 2852.754 (23968.831)\\
& & WLASSO & 1.292 (0.055) & 1.290 (0.055)  & 1.289 (0.055) & 0.208 (0.065) & 0.207 (0.065) & 0.206 (0.065)\\
& 1.00 & WMCEN & $\bm{1.237}$ $\bm{(0.049)}$ & $\bm{1.237 (0.049)}$ & $\bm{1.240 (0.047)}$ & $\bm{0.153 (0.037)}$ & $\bm{0.152 (0.037)}$  & $\bm{0.153 (0.036)}$ \\
&  & MCEN & 2.594 (0.655) & 2.586 (0.664) & 2.586 (0.665) & 28.649 (176.912) & 27.597 (181.941) & 27.359 (183.175)\\
&  & SEN & 2.857 (0.711) & 2.883 (0.708) & 2.923 (0.711) & 2844.378 (23964.838) & 2855.322 (23956.397) & 2858.504 (23940.471)\\
&  & WLASSO & 1.294 (0.056) & 1.292 (0.055) & 1.297 (0.057) & 0.198 (0.056) & 0.195 (0.055) & 0.201 (0.058)\\\hline
&   & $\xi$ & 0.02 & 0.05 & 0.10 & 0.02 & 0.05 & 0.10\\
$p$ & $\eta$ &  &  &  &  &  &  & \\\hline

100 & 0.25 & WMCEN & $\bm{1.377 (0.043)}$ & $\bm{1.378 (0.043)}$ & $\bm{1.401 (0.045)}$ & $\bm{0.011 (0.001)}$ & $\bm{0.011 (0.001)}$ & $\bm{0.012 (0.002)}$ \\
&  & MCEN & 2.473 (0.542) & 2.473 (0.540) & 2.472 (0.535) & 1.479 (9.455) & 1.512 (9.630) & 1.461 (9.968) \\
& & SEN & 2.393 (0.643) & 2.638 (0.789) & 2.425 (0.669) & 582.884 (5031.615) & 670.449 (5871.294) & 535.364 (4371.789) \\
&  & WLASSO & 3.202 (0.401) & 3.202 (0.400) & 3.206 (0.399) & 0.218 (0.144) & 0.218 (0.144) & 0.219 (0.144) \\
 & 0.50 & WMCEN &  $\bm{1.654 (0.057)}$ &  $\bm{1.640 (0.058)}$ &  $\bm{1.639 (0.062)}$ &  $\bm{0.026 (0.003)}$ &  $\bm{0.025 (0.003)}$ &  $\bm{0.024 (0.003)}$ \\
&  & MCEN & 3.514 (0.890) & 2.895 (0.545) & 2.901 (0.549) & 39.493 (243.653) & 1.099 (6.496) & 0.833 (5.079) \\
&  & SEN & 3.247 (0.812) & 3.355 (0.893) & 3.029 (0.726) & 730.828 (6363.460) & 608.572 (5044.895) & 713.699 (6253.216) \\
& & WLASSO & 3.507 (0.422) & 3.383 (0.397) & 3.382 (0.394) & 0.303 (0.173) & 0.248 (0.143) & 0.249 (0.143) \\
& 0.75 & WMCEN &  $\bm{1.916 (0.111)}$ &  $\bm{1.971 (0.123)}$ &  $\bm{1.930 (0.110)}$ &  $\bm{0.041 (0.005)}$ &  $\bm{0.045 (0.005)}$ &  $\bm{0.042 (0.004)}$ \\
& & MCEN & 3.565 (0.798) & 3.433 (0.722) & 3.556 (0.791) & 25.040 (139.366) & 18.090 (97.046) & 24.118 (133.008) \\
&  & SEN & 3.776 (0.838) & 3.756 (0.894) & 3.769 (0.889) & 666.696 (5723.088) & 631.024 (5330.103) & 723.491 (6093.913) \\
&  & WLASSO & 3.978 (0.484) & 3.979 (0.483) & 3.980 (0.484) & 61.319 (510.994) & 61.317 (510.980) & 61.315 (510.968) \\
& 1.00 & WMCEN & $\bm{2.217 (0.104)}$ & $\bm{2.261 (0.207)}$ & $\bm{2.253 (0.202)}$ & $\bm{0.070 (0.008)}$& $\bm{0.066 (0.009)}$ & $\bm{0.067 (0.009)}$ \\
&  & MCEN & 3.807 (0.754) & 3.789 (0.741) & 3.777 (0.763) & 23.917 (123.683) & 22.480 (114.956) & 20.724 (105.953) \\
&  & SEN & 4.413 (0.924) & 4.428 (1.017) & 4.351 (0.947) & 612.964 (5383.766) & 646.733 (5487.569) & 688.426 (6132.423) \\
&  & WLASSO & 4.248 (0.506) & 4.169 (0.482) & 4.063 (0.463) & 61.386 (511.104) & 0.686 (0.931) & 0.405 (0.175) \\\hline
\end{tabular}
}
\end{center}
$\quad \quad \quad \quad \quad \quad \quad$ 
\end{table}

\begin{table}
\begin{center}
\caption{Result of bias and Spearman's correlation coefficient of $\bm{B}$ in $p=12$ and $p=100$ for Error $1$. The bias values were too small in all methods; therefore, the results were shown up to the fourth decimal places.}\label{bico1}
\scalebox{0.8}{
\begin{tabular}{ccr rrr rrr} 
\toprule
& \quad  & & \multicolumn{3}{c}{bias of $\bm{\beta}_s$} &  \multicolumn{3}{c}{correlation of $\bm{B}$ }  \\\cmidrule(lr){4-6} \cmidrule(lr){7-9}
& & $\xi$ & 0.02 & 0.05 & 0.10 & 0.02 & 0.05 & 0.10\\
$p$ & $\eta$ & & & & & & &\\ 
\midrule
12 & 0.25 & WMCEN & -0.0031 &  $\bm{-0.0031}$ & $\bm{-0.0032}$ & $\bm{
0.401}$ & $\bm{0.407}$ & $\bm{0.415}$\\
& & MCEN & $\bm{-0.0030}$ &  -0.0048 & -0.0055 & 0.339 & 0.337 & 0.291\\
& & SEN & -0.0076 & -0.0089 & -0.0094 & 0.382 & 0.407 & $\bm{0.421}$\\
& & WLASSO & -0.0034 & -0.0035 & -0.0038 & 0.370 &  0.378 & 0.396\\
&  0.50 & WMCEN & $\bm{-0.0043}$ & $\bm{-0.0042}$ & $\bm{-0.0045}$ & 0.658  & 0.652 & 0.657 \\
& & MCEN & -0.0073 & -0.0074 & -0.0084 & $\bm{0.722}$ & $\bm{0.725}$ & $\bm{0.686}$\\
& & SEN & -0.0086 & -0.0087 & -0.0094 & 0.649 & 0.658 & 0.661 \\
& & WLASSO & -0.0052 & -0.0047 & -0.0045 & 0.640 & 0.639 & 0.646\\
&  0.75 & WMCEN & $\bm{-0.0045}$ & $\bm{-0.0045}$  & $\bm{-0.0045}$ & 0.766 & 0.770 & 0.772\\
& & MCEN & -0.0080 & -0.0121 & -0.0122 & $\bm{-0.814}$ & $\bm{-0.826}$ & $\bm{-0.831}$\\
& & SEN & -0.0073 & -0.0086 & -0.0088 & 0.767 & 0.774 & 0.789\\
& & WLASSO & -0.0047 & -0.0047  & -0.0047 &  0.764 & 0.768 & 0.773\\
& 1.00 & WMCEN & -0.0046 & $\bm{-0.0046}$ & $\bm{-0.0046}$ & 0.793 & 0.797  & 0.803 \\
&  & MCEN & -0.0167 & -0.0113 & -0.0114  & $\bm{0.831}$ & $\bm{0.831}$ & $\bm{0.837}$\\
&  & SEN & $\bm{-0.0037}$ & -0.0073 & -0.0053 & 0.802 & 0.811 & 0.811\\
&  & WLASSO & -0.0048 & -0.0048 & -0.0048 & 0.805 & 0.808 & 0.813\\\hline
&   & $\xi$ & 0.02 & 0.05 & 0.10 & 0.02 & 0.05 & 0.10\\
$p$ & $\eta$ &  &  &  &  &  &  & \\\hline

100 & 0.25 & WMCEN & -0.0106 & -0.0106 & -0.0106 & 0.370 & 0.372 & 0.365\\
&  & MCEN & -0.0098 & -0.0099 & -0.0095 & $\bm{0.393}$ & $\bm{0.391}$ & $\bm{0.460}$ \\
& & SEN & -0.0135 & -0.0146 & -0.0140 & 0.327 & 0.382 & 0.371 \\
&  & WLASSO & $\bm{-0.0073}$ & $\bm{-0.0073}$ & $\bm{-0.0072}$ & 0.279 & 0.279 & 0.277 \\
& 0.50 & WMCEN & -0.0168 & -0.0153 & -0.0150 &  0.468 &  0.460 & 0.457\\
&  & MCEN & -0.0130 & -0.0120 & -0.0131 & $\bm{0.517}$ & $\bm{0.540}$ & $\bm{0.537}$ \\
&  & SEN & -0.0187 & -0.0192 & -0.0216 & 0.505 & 0.507 & 0.480 \\
& & WLASSO & $\bm{-0.0116}$  & $\bm{-0.0117}$ & $\bm{-0.0116}$ & 0.428 & 0.429 & 0.428 \\
& 0.75 & WMCEN &  -0.0188 &  -0.0187 &  -0.0182 &  0.498 &  0.498 &  0.497 \\
& & MCEN & -0.0173 & -0.0173 & -0.0173 & 0.521 & 0.521 & 0.520\\
&  & SEN & -0.0227 & -0.0225 & -0.0231 & $\bm{0.588}$ & $\bm{0.562}$  & $\bm{0.558}$  \\
&  & WLASSO & $\bm{-0.0138}$ & $\bm{-0.0138}$ & $\bm{-0.0138}$ & 0.505 & 0.504 & 0.504 \\
& 1.00 & WMCEN & -0.0174 & -0.0165 & -0.0200 & 0.509 & 0.509 & 0.512 \\
&  & MCEN & -0.0194 & -0.0195 & -0.0195 & 0.537 & 0.537 & 0.537\\
&  & SEN & -0.0231 & -0.0265 & -0.0233 & $\bm{0.605}$ & $\bm{0.608}$ & $\bm{0.589}$ \\
&  & WLASSO & $\bm{-0.0147}$ & $\bm{-0.0148}$ & $\bm{-0.0147}$ & 0.538 & 0.540 & 0.538 \\\hline
\end{tabular}
}
\end{center}
$\quad \quad \quad \quad \quad \quad \quad$ 
\end{table}

\begin{table}
\begin{center}
\caption{Result of bias and Spearman's correlation coefficient of $\bm{B}$ in  $p=12$ and $p=100$ for Error $2$. The bias values were too small in all methods; therefore, the results were shown up to the fourth decimal places. }\label{bico2}
\scalebox{0.8}{
\begin{tabular}{ccr rrr rrr} 
\toprule
& \quad  & & \multicolumn{3}{c}{bias of $\bm{\beta}_s$} &  \multicolumn{3}{c}{correlation of $\bm{B}$ }  \\\cmidrule(lr){4-6} \cmidrule(lr){7-9}
& & $\xi$ & 0.02 & 0.05 & 0.10 & 0.02 & 0.05 & 0.10\\
$p$ & $\eta$ & & & & & & &\\ 
\midrule
12 & 0.25 & WMCEN & -0.0040 & -0.0040 & -0.0040 & $\bm{0.345}$ & $\bm{0.354}$ & $\bm{0.364}$\\
& & MCEN & $\bm{-0.0001}$ &  -0.0116 & -0.0056 & 0.044 & 0.059 & 0.039\\
& & SEN & -0.0277 & -0.0190 & -0.0216 & 0.034 & 0.040 & 0.038\\
& & WLASSO & -0.0033 & $\bm{-0.0032}$  & $\bm{-0.0032}$ &  0.324 & 0.334 & 0.345\\
&  0.50 & WMCEN & -0.0057 & -0.0056 & -0.0057 & $\bm{0.586}$   &  $\bm{0.592}$ & $\bm{0.597}$ \\
& & MCEN & -0.0316 & -0.0301 & -0.0278 & 0.095 & 0.095 & 0.101\\
& & SEN & -0.0281 & -0.0233 & -0.0290 & 0.077 & 0.083 & 0.083 \\
& & WLASSO & $\bm{-0.0036}$ & $\bm{-0.0036}$  & $\bm{-0.0036}$ & 0.568 & 0.573 & 0.580\\
&  0.75 & WMCEN & $\bm{-0.0040}$ & $\bm{-0.0009}$ & -0.0064 & $\bm{0.720}$ & 0.706 & $\bm{0.724}$\\
& & MCEN & -0.0224 & -0.0229 & -0.0220 & 0.121 & 0.126 &  0.128\\
& & SEN & -0.0414 & -0.0364 & -0.0382 & 0.103 & 0.108 & 0.117\\
& & WLASSO & -0.0043 & -0.0043 & $\bm{-0.0043}$ & 0.712 & $\bm{0.716}$ & 0.721\\
& 1.00 & WMCEN & -0.0062 & $\bm{-0.0041}$ & -0.0064 & $\bm{0.777}$ & $\bm{0.778}$ & $\bm{0.782}$\\
&  & MCEN & -0.0349 & -0.0347 & -0.0340 & 0.149 & 0.151 & 0.153\\
&  & SEN & -0.0310 & -0.0285 & -0.0374 & 0.143 & 0.147 & 0.138\\
&  & WLASSO & $\bm{-0.0044}$ & -0.0044 & $\bm{-0.0044}$ & 0.774 & 0.777 & 0.782\\\hline
&   & $\xi$ & 0.02 & 0.05 & 0.10 & 0.02 & 0.05 & 0.10\\
$p$ & $\eta$ &  &  &  &  &  &  & \\\hline

100 & 0.25 & WMCEN & -0.0148 & -0.0125  & -0.0147  & $\bm{0.374}$ & $\bm{0.359}$ & $\bm{0.365}$ \\
&  & MCEN & $\bm{-0.0063}$  & $\bm{-0.0073}$ & -0.0075 & 0.079 & 0.055 & 0.081 \\
& & SEN & -0.0100 & -0.0122 & -0.0102 & 0.065 & 0.061 & 0.063 \\
&  & WLASSO & -0.0075 & -0.0075 & $\bm{-0.0074}$ & 0.205 & 0.204 & 0.204\\
& 0.50 & WMCEN &  -0.0197 & -0.0193 & -0.0193 &  $\bm{0.442}$ &  $\bm{0.441}$ & $\bm{0.439}$ \\
&  & MCEN & -0.0151 & -0.0151 & -0.0156 & 0.101 & 0.101 & 0.101 \\
&  & SEN & -0.0206 & -0.0172 & -0.0194 & 0.107 & 0.101 & 0.102 \\
& & WLASSO & $\bm{-0.0127}$  & $\bm{-0.0142}$ & $\bm{-0.0143}$ & 0.310 & 0.301 & 0.297\\
& 0.75 & WMCEN &  -0.0228 &  -0.0272 &  -0.0243 &  $\bm{0.472}$ & $\bm{0.476}$ & $\bm{0.479}$\\
& & MCEN & $\bm{-0.0205}$ & $\bm{-0.0207}$ & -0.0207 & 0.092 & 0.092 & 0.092 \\
&  & SEN & -0.0279 & -0.0283 & -0.0262 & 0.130 & 0.137 & 0.130 \\
&  & WLASSO & -0.0211 & $\bm{-0.0212}$ & $\bm{-0.0196}$ & 0.358 & 0.358 & 0.369 \\
& 1.00 & WMCEN & $\bm{-0.0230}$ & $\bm{-0.0222}$ & $\bm{-0.0223}$ & $\bm{0.472}$ & $\bm{0.472}$ & $\bm{0.472}$\\
&  & MCEN & -0.0269 & -0.0269 & -0.0267 & 0.104 & 0.104  & 0.104  \\
&  & SEN & -0.0358 & -0.0350 & -0.0356 & 0.160 & 0.156 & 0.164 \\
&  & WLASSO & -0.0250 & -0.0250 & -0.0251 & 0.399 & 0.399 & 0.399 \\\hline
\end{tabular}
}
\end{center}
$\quad \quad \quad \quad \quad \quad \quad$ 
\end{table}

\begin{table}
\begin{center}
\caption{Result of bias and Spearman's correlation coefficient of $\bm{B}$ in  $p=12$ and $p=100$ for Error $3$. The bias values were too small in all methods; therefore, the results were shown up to the fourth decimal places. }\label{bico3}
\scalebox{0.8}{
\begin{tabular}{ccr rrr rrr} 
\toprule
& \quad  & & \multicolumn{3}{c}{bias of $\bm{\beta}_s$} &  \multicolumn{3}{c}{correlation of $\bm{B}$ }  \\\cmidrule(lr){4-6} \cmidrule(lr){7-9}
& & $\xi$ & 0.02 & 0.05 & 0.10 & 0.02 & 0.05 & 0.10\\
$p$ & $\eta$ & & & & & & &\\ 
\midrule
12 & 0.25 & WMCEN & $\bm{-0.0046}$ & $\bm{-0.0045}$ & $\bm{-0.0046}$ & $\bm{0.248}$ & $\bm{0.256}$ & $\bm{0.264}$\\
& & MCEN & -0.0077 & -0.0120 & -0.0128 & 0.189 & 0.201 & 0.216\\
& & SEN & -0.0168 & -0.0148 & -0.0149 & 0.188 & 0.197 & 0.248\\
& & WLASSO & -0.0048 & -0.0048 & -0.0049 & 0.227 & 0.236 & 0.248\\
&  0.50 & WMCEN & -0.0046 & -0.0071 & -0.0072 & $\bm{0.450}$ & $\bm{0.453}$ & $\bm{0.459}$ \\
& & MCEN &  -0.0163 & -0.0112 & -0.0175 & 0.393 & 0.388 & 0.403\\
& & SEN & -0.0161 & -0.0184 & -0.0167 & 0.392 & 0.399 & 0.415 \\
& & WLASSO & $\bm{-0.0056}$ & $\bm{-0.0054}$  & $\bm{-0.0053}$ & 0.435 & 0.441 & 0.449\\
&  0.75 & WMCEN & -0.0082 & -0.0083 & -0.0081 & $\bm{0.604}$ & $\bm{0.606}$ & $\bm{0.608}$\\
& & MCEN & -0.0183 & -0.0185 & -0.0190 & 0.558 & 0.561 &  0.568\\
& & SEN & -0.0188 & -0.0148 & -0.0201 & 0.539 & 0.544 & 0.559\\
& & WLASSO & $\bm{-0.0069}$ & $\bm{-0.0069}$  & $\bm{-0.0065}$ & 0.593 & 0.597 & 0.602\\
& 1.00 & WMCEN & -0.0086 & -0.0085 & -0.0086 & 0.707 & 0.704  & 0.707 \\
&  & MCEN & -0.0151 & -0.0232 & -0.0234 & $\bm{0.731}$ & $\bm{0.722}$ & $\bm{0.725}$\\
&  & SEN & -0.0186 & -0.0166 & -0.0149 & 0.657 & 0.654 & 0.662\\
&  & WLASSO & $\bm{-0.0062}$ & $\bm{-0.0062}$ & $\bm{-0.0060}$ & 0.688 & 0.690 & 0.694\\\hline
&   & $\xi$ & 0.02 & 0.05 & 0.10 & 0.02 & 0.05 & 0.10\\
$p$ & $\eta$ &  &  &  &  &  &  & \\\hline

100 & 0.25 & WMCEN & -0.0135 & -0.0137  & -0.0135  & $\bm{0.303}$ & $\bm{0.304}$ & 0.300 \\
&  & MCEN & -0.0139 & -0.0148 & -0.0154 & 0.287 & 0.295 & $\bm{0.307}$ \\
& & SEN & -0.0164 & -0.0158 & -0.0148 & 0.253 & 0.249 & 0.249 \\
&  & WLASSO & $\bm{-0.0069}$ & $\bm{-0.0071}$ & $\bm{-0.0071}$ & 0.169 & 0.169 & 0.169 \\
& 0.50 & WMCEN &  -0.0244 & -0.0248 & -0.0236 &  0.406 &  0.407 & 0.407 \\
&  & MCEN & -0.0186 & -0.0192 & -0.0192 & $\bm{0.492}$ & $\bm{0.495}$ & $\bm{0.478}$ \\
&  & SEN & -0.0268 & -0.0258 & -0.0236 & 0.383 & 0.397 & 0.403 \\
& & WLASSO & $\bm{-0.0136}$  & $\bm{-0.0135}$ & $\bm{-0.0137}$ & 0.293 & 0.290 & 0.295 \\
& 0.75 & WMCEN &  -0.0275 & -0.0277 & -0.0276 &  0.453 &  0.455 & 0.456 \\
& & MCEN & -0.0222 & -0.0231 & -0.0228 & $\bm{0.556}$ & $\bm{0.551}$ & $\bm{0.553}$ \\
&  & SEN & -0.0321 & -0.0294 & -0.0330 & 0.498 & 0.498 & 0.496 \\
&  & WLASSO & $\bm{-0.0187}$ & $\bm{-0.0190}$ & $\bm{-0.0185}$ & 0.375 & 0.377 & 0.374  \\
& 1.00 & WMCEN & -0.0277 & -0.0282 & -0.0280 & 0.478 & 0.478 & 0.478 \\
&  & MCEN & -0.0257 & -0.0257 & -0.0253 & $\bm{0.555}$ & $\bm{0.555}$ & $\bm{0.540}$ \\
&  & SEN & -0.0368 & -0.0354 & -0.0355 & 0.547 & 0.540  & 0.530 \\
&  & WLASSO & $\bm{-0.0218}$ & $\bm{-0.0219}$ & $\bm{-0.0219}$ & 0.434 & 0.437 & 0.434 \\\hline
\end{tabular}
}
\end{center}
$\quad \quad \quad \quad \quad \quad \quad$ 
\end{table}

\begin{table}
\begin{center}
\caption{Result of bias and Spearman's correlation coefficient of $\bm{B}$ in  $p=12$ and $p=100$ for Error $4$. The bias values were too small in all methods; therefore, the results were shown up to the fourth decimal places. }\label{bico4}
\scalebox{0.8}{
\begin{tabular}{ccr rrr rrr} 
\toprule
& \quad  & & \multicolumn{3}{c}{bias of $\bm{\beta}_s$} &  \multicolumn{3}{c}{correlation of $\bm{B}$ }  \\\cmidrule(lr){4-6} \cmidrule(lr){7-9}
& & $\xi$ & 0.02 & 0.05 & 0.10 & 0.02 & 0.05 & 0.10\\
$p$ & $\eta$ & & & & & & &\\ 
\midrule
12 & 0.25 & WMCEN & -0.0019 & -0.0042 & -0.0042 & $\bm{0.232}$ & $\bm{0.244}$ & $\bm{0.252}$\\
& & MCEN & 0.0332 & 0.0249 & 0.0062 & 0.056 & 0.059 & 0.064\\
& & SEN & $\bm{0.0018}$ & 0.0060 & $\bm{0.0033}$  & 0.037 & 0.045 & 0.045\\
& & WLASSO & -0.0039 & $\bm{-0.0040}$ & -0.0047 & 0.208 & 0.213 & 0.218\\
&  0.50 & WMCEN & -0.0072 & -0.0072 & -0.0072 & $\bm{0.404}$   &  $\bm{0.409}$ & $\bm{0.415}$ \\
& & MCEN & 0.0319 & 0.0301 & 0.0239 & 0.099 & 0.101 & 0.108\\
& & SEN & -0.0247 & -0.0299 & -0.0107 & 0.078 & 0.074 & 0.090 \\
& & WLASSO & $\bm{-0.0059}$ & $\bm{-0.0059}$  & $\bm{-0.0056}$ & 0.368 & 0.373 & 0.381)\\
&  0.75 & WMCEN & $\bm{-0.0051}$ & -0.0090 & -0.0091 & $\bm{0.530}$ & $\bm{0.538}$ & $\bm{0.542}$\\
& & MCEN & 0.0140 & 0.0127 & 0.0009 & 0.154 & 0.155 &  0.155\\
& & SEN & -0.0319 & -0.0272 & -0.0293 & 0.118 & 0.125 & 0.126\\
& & WLASSO & -0.0064 & $\bm{-0.0066}$  & $\bm{-0.0068}$ & 0.503 & 0.507 & 0.512\\
& 1.00 & WMCEN & $\bm{-0.0060}$ & $\bm{-0.0060}$ & $\bm{-0.0059}$ & $\bm{0.632}$ & $\bm{0.636}$  & $\bm{0.637}$ \\
&  & MCEN & 0.0587 & 0.0620 & 0.0611 & 0.189 & 0.185 & 0.190\\
&  & SEN & -0.0387 & -0.0381 & -0.0468 & 0.161 & 0.167 & 0.166\\
&  & WLASSO & $\bm{-0.0077}$ & $\bm{-0.0081}$ & $\bm{-0.0072}$ & 0.619 & 0.621 & 0.625\\\hline
&   & $\xi$ & 0.02 & 0.05 & 0.10 & 0.02 & 0.05 & 0.10\\
$p$ & $\eta$ &  &  &  &  &  &  & \\\hline

100 & 0.25 & WMCEN & -0.0142 & -0.0142  & -0.0138  & $\bm{0.266}$ & $\bm{0.266}$ & $\bm{0.260}$\\
&  & MCEN & -0.0230 & -0.0229 & -0.0228 & 0.085 & 0.086 & 0.088 \\
& & SEN & 0.1469 & 0.1807 & 0.1135 & 0.071 & 0.056 & 0.079 \\
&  & WLASSO & $\bm{-0.0087}$ & $\bm{-0.0087}$ & $\bm{-0.0085}$ & 0.106 & 0.106 & 0.106 \\
& 0.50 & WMCEN & -0.0257 & -0.0260 & -0.0266 &  $\bm{0.360}$ & $\bm{0.362}$ & $\bm{0.363}$ \\
&  & MCEN & -0.0248 & -0.0444 & -0.0445 & 0.155 & 0.161 & 0.164 \\
&  & SEN & 0.1816 & 0.1224 & 0.1638 & 0.119 & 0.099 & 0.115 \\
& & WLASSO & $\bm{-0.01162}$  & $\bm{-0.01161}$ & $\bm{-0.01162}$ & 0.178 & 0.185 & 0.184 \\
& 0.75 & WMCEN & -0.0381 & -0.0343 & -0.0383 &  $\bm{0.413}$ & $\bm{0.405}$ & $\bm{0.412}$ \\
& & MCEN & -0.0315 & -0.0366 & -0.0338 & 0.227 & 0.234 & 0.228 \\
&  & SEN & 0.1379 & 0.1299 & 0.1457 & 0.127 & 0.157 & 0.142 \\
&  & WLASSO & $\bm{-0.0021}$ & $\bm{-0.0020}$ & $\bm{-0.0019}$ & 0.224 & 0.225 & 0.225 \\
& 1.00 & WMCEN & -0.0408 & -0.0442 & -0.0442 & $\bm{0.430}$ & $\bm{0.435}$ & $\bm{0.434}$ \\
&  & MCEN & -0.0194 & -0.0180 & -0.0175 & 0.285 & 0.287 & 0.289 \\
&  & SEN & 0.1111 & 0.1191 & 0.1372 & 0.177 & 0.155 & 0.195 \\
&  & WLASSO & $\bm{-0.0084}$ & $\bm{-0.0301}$ & $\bm{-0.0298}$ & 0.268 & 0.273 & 0.280 \\\hline
\end{tabular}
}
\end{center}
$\quad \quad \quad \quad \quad \quad \quad$ 
\end{table}
%

\begin{figure}[htbp]
\begin{center}
\includegraphics[scale=0.68]{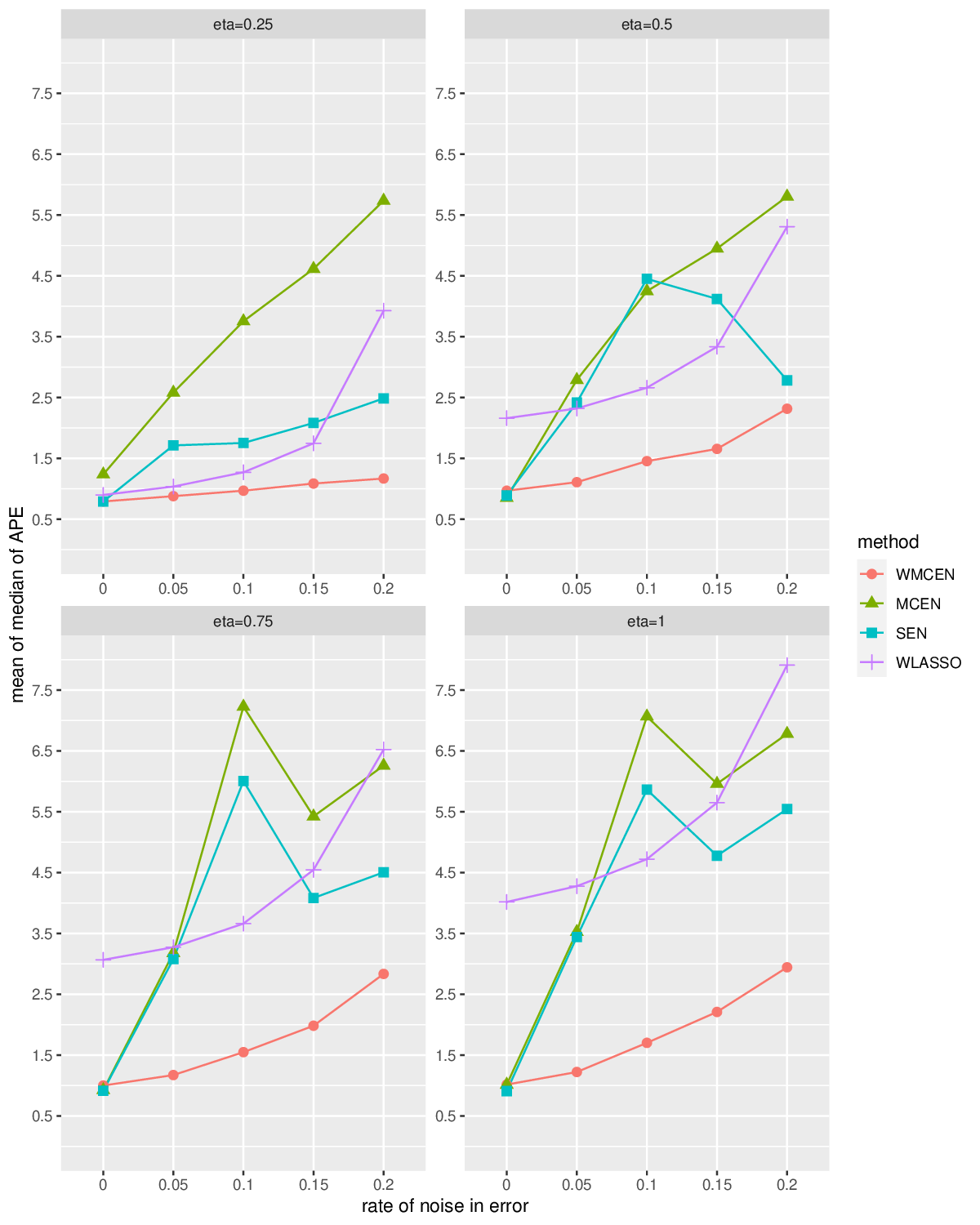}
\caption{Results of the mean of the median APE by $\eta$. The vertical axis indicates the mean of the median APE, and the horizontal axis indicates proportion of contaminated distribution.}
\label{api_p50}
\end{center}
\end{figure}
%

\begin{figure}[htbp]
\begin{center}
\includegraphics[scale=0.65]{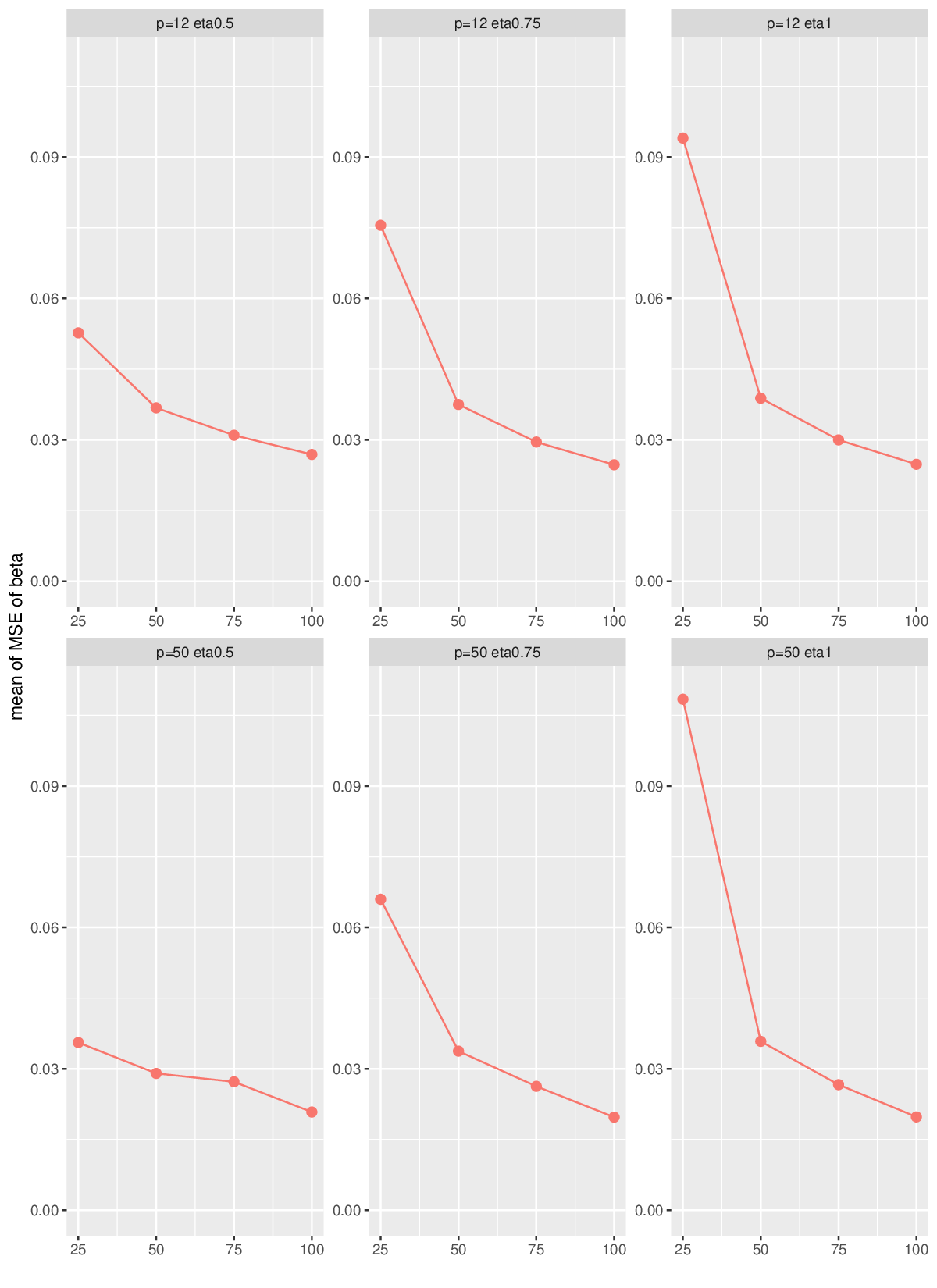}
\caption{Results of MSE of $\bm{\beta}_s$ by number of response variable $p$ and $\eta$. The vertical axis indicates the mean of MSE of $\bm{\beta}_s$, and the horizontal axis is the number of subject.}
\label{asym}
\end{center}
\end{figure}

Next, by observing in the covariate variables of Factor 1.2 on the median of APE, the proposed method was superior to SEN in $\eta=0.50, 0.75,$ and $1.00$ in $p=12$ for Error 1. However, the values of MCEN and SEN were smaller than those in the proposed method, in the same pattern for $p=100$. 
In Error 3 for $p=12$, the proposed method outperformed MCEN for most patterns of both evaluation measures, while MCEN was better for $p=100$. 
In Error 1 and Error 3, the values of the proposed method and MCEN were close to each other, regardless of the value of the covariate variables. 
The results on Error 2 and Error 4 did not vary depending on the covariate variables.  

As for Factor 1.3, parameter $\eta$, the values of all methods increased in proportion to the augment of $\eta$. Meanwhile, the values did not tend to change depending on $\xi$, Factor 1.4.

Now we see the results of the error distribution, Factor 1.5. In Error 1, the proposed method was almost stable for $p=12$ in the median APE, while SEN and MCEN were better than the proposed method for $p=100$. MCEN had the smallest MSE of ${\bm \beta}_s$ for both $p=12$ and $p=100$. However, the differences in values among the methods were small in both the median APE and the MSE of ${\bm \beta}_s$. 
In the case of Error 2 where the error contains outliers, for both $p=12$ and $p=100$, the proposed method was better than the other compared methods in both the median APE and the MSE of ${\bm \beta}_s$. Focusing on the differences, the differences in values between the proposed method and WLASSO were smaller than those between the proposed method and MCEN and SEN.
In Error 3 with the error distribution following $t$ distribution, the median APE of the proposed method was better in $(\eta, \xi) = (0.25, 0.05), (0.25, 0.10)$, and $(0.50, 0.10)$, and MCEN had better results in the other pattern of $\eta$ and $\xi$. In the MSE of ${\bm \beta}_s$, the proposed method had the smallest values for all for $p=12$, except $\eta = 0.25$. For $p=100$, MCEN had smaller values than the proposed method; however, the differences were minor.
Finally, in Error 4, which is another heavy-tailed error distribution, the proposed method was better than all the other compared methods. 
For $p=12$ in Error 4, WLASSO was also stable in terms of both the median of APE and MSE of ${\bm \beta}_s$ compared to MCEN and SEN.

Next, we consider the results of the bias of $\bm{\beta}_s$ and the correlation of $\bm{B}$, as shown in Table \ref{bico1} to Table \ref{bico4}. 
Regarding the bias of $\bm{\beta}_s$, there were no large differences among all methods, and the bias values were small in all patterns. Specifically, the values of all methods were close to $0$ in $p=12$.
In the correlation, all patterns showed that the correlation was higher with increasing values of $\eta$, with no differences according to $\xi$.
The proposed method was better than the other methods in Error 1 for $p=12$ (Table \ref{bico1}), Error 2 (Table \ref{bico2}), Error 3 except $\xi=1.00$ for $p=12$ and that in $\eta = 0.25$ for $p=100$  (Table \ref{bico3}), and Error 4 (Table \ref{bico1}). 
The values of MCEN and SEN were greater than those of the proposed method in Error 1 except $\eta = 0.25$, while the results of Error 3 were closer to those of the proposed method. Additionally, the values in Error 2 and Error 4 were low compared to those of the proposed method.

The results for the proportions of the contaminated distribution in simulation 2 are shown in Figure \ref{api_p50}, which is plotted by $\eta$. Figure \ref{api_p50} shows that the mean values of the proposed method are lower than those of the compared method in the presence of noise. Even when the proportion of noise in the error distribution increases, the trend of the values is not large compared to that of the compared methods. In this simulation results, MCEN and SEN showed particularly unstable estimation in the presence of noise.

Finally, the results of the performance of estimators in simulation 3 are plotted in Figure \ref{asym}. The MSE of the difference between the predicted coefficients vector $\hat{\bm{\beta}}_s$ and true coefficients $\bm{\beta}^*_s$ were found to decrease, depending on the incrase in the number of the subject in all patterns.

\subsection{Discussion of simulation results}

In the simulation 1, the proposed method was superior when the errors followed heavy-tailed distributions or the data contained outliers, as in Error $2$, Error $3$, and Error $4$. 
Specifically, in Error $2$, which contained outliers, and Error $4$, which had a strong heavy-tailed setting, the proposed method performed much better than the other compared methods in both the median APE and MSE of ${\bm \beta}_s$. This indicates that the proposed method is stable to heavy-tailed error distributions and outliers in the error. 
Hence, the results obtained by our proposed method were similar results to those in \cite{wlasso}. 

In addition, among the compared methods, WLASSO had closer results to the proposed method than MCEN and SEN in Error 2 in simulation 1. 
This shows that a Wilcoxon-type regression function is effective when the data contain outliers. 
Moreover, our proposed method extended the Wilcoxon-type regression function in the framework of multivariate regression, which facilitated the consideration of the correlation of response variables by the penalty term while taking clustering into account.
We confirmed the notable differences between the proposed method and WLASSO, especially when increasing proportions of contaminated distribution in the simulation 2. 
That is, the $L_2$ penalty term of the proposed method contributes the stability of estimation in the situation with outlier.

We have discussed the case of outliers and heavy-tailed distributions for the proposed method. We now discuss the case with the absence of noise.
When compared to MCEN and SEN, in Error 1 in simulation 1, where the error follows normal distribution, the differences between the proposed method were not large.
This result is consistent with those in the error with normal distribution in \cite{wlasso}. 
Furthermore, in the simulation 3, we confirmed that the MSE of $\bm{\beta}_s$ decreased as the number of subject increased in the proposed method. The proposed method is expected to have good property based on empirical investigation.

Therefore, in these simulations, the proposed method demonstrated high estimation accuracy for outliers and heavy-tailed distributions and retained the same estimation accuracy as the compared methods, even in situations where the error distribution follows a normal distribution.\\

\section{Real Example}\label{real}

In this section, we applied the proposed method to a genetic dataset named "chin07" from package lol in R software \citep{chin07} to verify the method’s usefulness. The dataset consists of copy number patterns and mRNA expression levels in genomic regions with candidate oncogenes for breast cancer. Recently, with the development of computational technologies, considerable research has been conducted on identifying genomic regions with candidate oncogenes through genome-wide profiling, aiming to suppress the expression of cancer. 
Altered DNA copy numbers are considered one of the causes of genetic abnormalities leading to disease.
The key to identifying which DNA copy number alternation patterns affect the mRNA expression level from this dataset. Therefore, it is necessary to improve the accuracy of estimation along with variable selection. 
This dataset consists of a matrix of seven types of mRNA expressions highly relevant to breast cancer with $106$ samples and a matrix of DNA copy number data for 339 regions with $106$ samples. We set the mRNA expression levels as response variables and the DNA copy numbers as explanatory variables. 
Figure E.1 in Supplementary material draws a scatter plot, histogram and correlations of the response variables. 
GI$\_$17318566-A, Hs.500472-S, and GI$\_$4758297-S appear to have right-tailed distribution, while GI$\_$38505204-S has a left-tailed distribution.
GI$\_$17318560-A contains outliers.
We compare the proposed method with MCEN, SEN, and WLASSO as in the numerical simulation. Tuning parameters were determined by cross validation in the same way as in the simulation.

\begin{figure}[htbp]
\begin{center}
\includegraphics[scale=0.6]{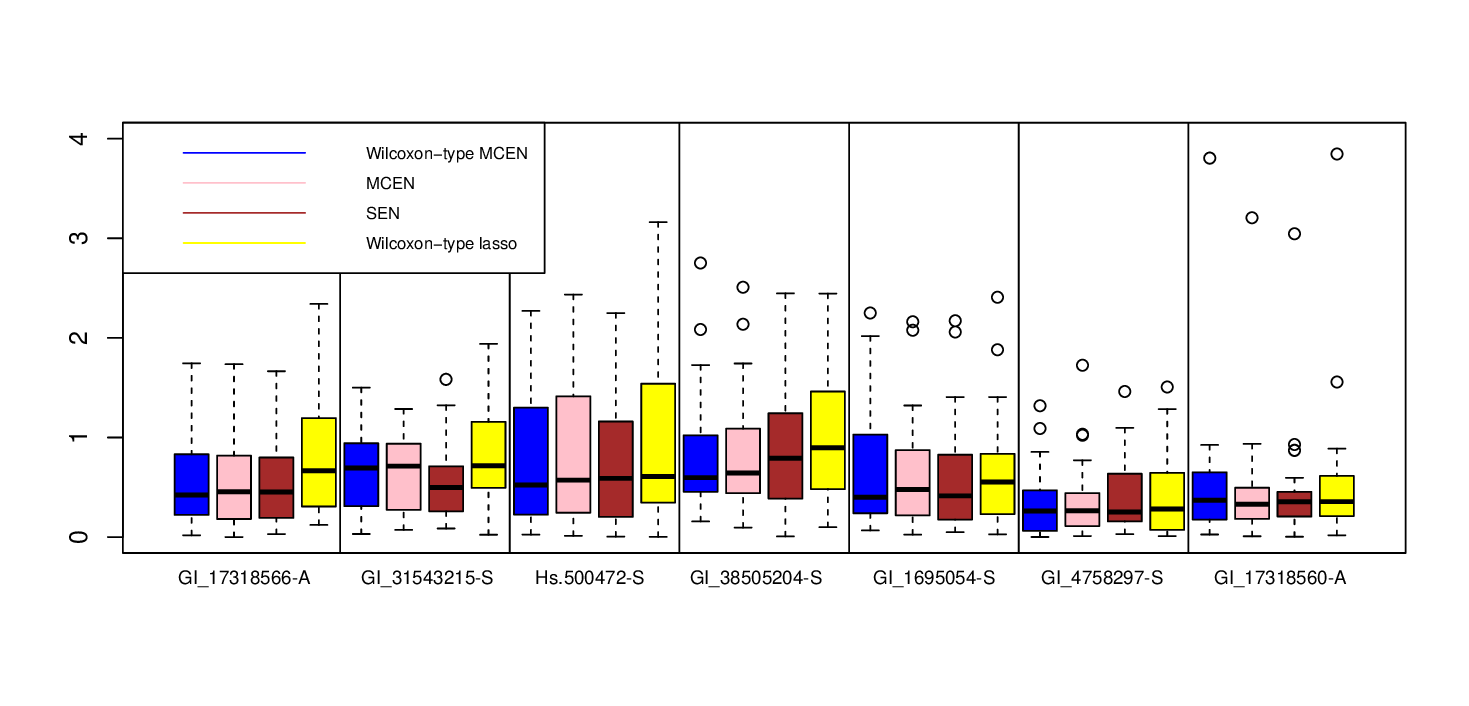}
\caption{Results of the real data for each response variable. The vertical axis indicates the APE for each response variable. The value of the center of the boxplot is the median APE.}
\label{resmeth}
\end{center}
\end{figure}

The results for the response variables are plotted in Figure \ref{resmeth}. The vertical axis describes the median APE. 
We compare the result of the proposed method to each compared method. 
First, compared with MCEN, the proposed method showed good results in the response variables GI$\_$17318566-A, GI$\_$31543215-S, Hs.500472-S, GI$\_$38505204-S, and GI$\_$16950654-S.
The proposed method also performed better than WLASSO for all response variables. 
In terms of the comparison with SEN, the proposed method had smaller evaluation values in the response variables GI$\_$17318566-A, Hs.500472-S, GI$\_$38505204-S, and GI$\_$17318560-S.

The results differed for each response variable; thus, we discuss the results from the distribution of the response variables in Figure E.1 in Supplementary material. 
The proposed method was superior to the other methods in the response variables with a heavy tail such as GI$\_$17318566-A, Hs.500472-S, GI$\_$38505204-S, and GI$\_$4758297-S. 
Furthermore, the proposed method was better than MCEN, regardless of the shape of the distribution of the response variables.
Next, compared with SEN, 
the proposed method had better outcomes for the variables GI$\_$17318566-A, Hs.500472-S and GI$\_$38505204-S, which followed heavy-tailed distributions.
In the variables GI$\_$31543215-S, GI$\_$4758297-S, and GI$\_$17318560-A, the proposed method was minimally inferior to SEN, which might be due to the selection of the tuning parameters on each response variable in SEN.

In real data application, the proposed method was better than the other methods in the responses GI$\_$17318566-A, Hs.500472-S, GI$\_$38505204-S, and GI$\_$4758297-S, which followed heavy-tailed distribution. The proposed method also showed the stability of heavy-tailed situations in real data. 
Meanwhile, in the variables GI$\_$4758297-S and GI$\_$17318560-A, the proposed method was minimally inferior to SEN.
The difference in the median APE between the proposed method and SEN was small. 
This may be because SEN selected the tuning parameters for each response variable, while the proposed method selected single tuning parameters for all response variables in this study.

\section{Discussion and Conclusion}

We showed the efficacy of the proposed method through the numerical simulation and real data application. 
From the results from numerical simulation and real data application, we infer that this method can provide high estimation accuracy when applied to actual multivariate data containing heavy-tailed distributions and outliers. 
To improve our proposed method, these five points need to be considered.
First, in this study, we set common tuning parameters for all response variables. 
We expect that the estimation accuracy of the method would be enhanced if different tuning parameters are set depending on each response variable. 
Second, the Wilcoxon-type regression is more efficient with respect to the tailed distribution or outliers of the response. However, it is not very robust against the outliers in explanatory variables. To overcome this drawback, the weight of the Wilcoxon-type regression can be set, as proposed in several studies \citep{naranjo1994bounded, sievers1983weighted, wwscad}. 
This will need further consideration in situations where the explanatory variables contain outliers. 
Third, theoretical analyses of robustness on the proposed method also needs to be considered. Specifically, the robustness of the proposed method to outliers needs to be examined. Fourth, as discussed in {\rm Remark 2} and {\rm Remark 3}, the asymptotic properties of multivariate Wilcoxon-type regression using $L_1$ norm with and without sparse penalty terms are guaranteed based on existing results \citep{chang1999high, heiler1988, wwscad, johnson2008rank}. For the property of the proposed method, which also contains clustering term (the third term of Eq. (\ref{WMCEN1})), we confirmed that the estimated coefficient matrix was converged to true coefficient matrix as increasing the number of subject through the numerical simulation 3. Although we investigated the property empirically, it needs to further examination in terms of asymptotic theory.
Fifth, our extension of the Wilcoxon-type regression to multivariate regression is formulated using $L_1$ norm, while the method proposed by \citet{multiWil} was extended using $L_2$ norm. 
By using the $L_1$ norm, the updated formula of the existing MCEN in combination with the MM algorithm can be employed. 
However, we need to compare the two methods from various perspectives including a theoretical analysis.
Finally, we extended the Wilcoxon-type regression in the framework of multivariate regression. 
However, several robust reduced-rank regressions for multivariate regression have also been proposed \citep{Chao2021, Ding2021, Wang2022, Tan2022, Mishra2022, Zhao2017, She2017} to deal with correlation of the response variables.
The difference between our proposed method and these robust reduced-rank methods is that the proposed method has a clustering term to group the fitted values of the response variables, which allows for consideration of the correlation among the response variables. This needs to be considered in future studies.


\bibliographystyle{unsrtnat}
\bibliography{references}  

\begin{thebibliography}{57}
\providecommand{\natexlab}[1]{#1}
\providecommand{\url}[1]{\texttt{#1}}
\expandafter\ifx\csname urlstyle\endcsname\relax
  \providecommand{\doi}[1]{doi: #1}\else
  \providecommand{\doi}{doi: \begingroup \urlstyle{rm}\Url}\fi

\bibitem[Breiman and Friedman(1997)]{breiman1997}
L.~Breiman and J.H. Friedman.
\newblock Predicting multivariate responses in multiple linear regression.
\newblock \emph{Journal of the Royal Statistical Society: Series B (Statistical
  Methodology)}, 59\penalty0 (1):\penalty0 3--54, 1997.

\bibitem[Rothman et~al.(2010)Rothman, Levina, and Zhu]{MRCE}
A.J. Rothman, E.~Levina, and J.~Zhu.
\newblock Sparse multivariate regression with covariance estimation.
\newblock \emph{Journal of Computational and Graphical Statistics}, 19\penalty0
  (4):\penalty0 947--962, 2010.

\bibitem[Peng et~al.(2010)Peng, Zhu, Bergamaschi, Han, Noh, Pollack, and
  Wang]{peng2010regularized}
J.~Peng, J.~Zhu, A.~Bergamaschi, W.~Han, D.Y. Noh, J.R. Pollack, and P.~Wang.
\newblock Regularized multivariate regression for identifying master predictors
  with application to integrative genomics study of breast cancer.
\newblock \emph{The annals of applied statistics}, 4\penalty0 (1):\penalty0
  53--77, 2010.

\bibitem[Kim and Xing(2012)]{kim2012tree}
S.~Kim and E.P. Xing.
\newblock Tree-guided group lasso for multi-response regression with structured
  sparsity, with an application to e{QTL} mapping.
\newblock \emph{The Annals of Applied Statistics}, 6\penalty0 (3):\penalty0
  1095--1117, 2012.

\bibitem[Chen et~al.(2017)Chen, Iyengar, and Iyengar]{chen2017modeling}
Y.~Chen, R.~Iyengar, and G.~Iyengar.
\newblock Modeling multimodal continuous heterogeneity in conjoint analysis—a
  sparse learning approach.
\newblock \emph{Marketing Science}, 36\penalty0 (1):\penalty0 140--156, 2017.

\bibitem[Cook et~al.(2010)Cook, Li, and Chiaromonte]{cook2010envelope}
R.D. Cook, B.~Li, and F.~Chiaromonte.
\newblock Envelope models for parsimonious and efficient multivariate linear
  regression.
\newblock \emph{Statistica Sinica}, 20\penalty0 (3):\penalty0 927--960, 2010.

\bibitem[Cook and Zhang(2015)]{cook2015foundations}
R.D. Cook and X.~Zhang.
\newblock Foundations for envelope models and methods.
\newblock \emph{Journal of the American Statistical Association}, 110\penalty0
  (510):\penalty0 599--611, 2015.

\bibitem[Sun et~al.(2015)Sun, Zhu, Liu, and Ibrahim]{sun2015sprem}
Q.~Sun, H.~Zhu, Y.~Liu, and J.G. Ibrahim.
\newblock Sprem: sparse projection regression model for high-dimensional linear
  regression.
\newblock \emph{Journal of the American Statistical Association}, 110\penalty0
  (509):\penalty0 289--302, 2015.

\bibitem[Price and Sherwood(2018)]{MCEN}
B.~S. Price and B.~Sherwood.
\newblock A cluster elastic net for multivariate regression.
\newblock \emph{Journal of Machine Learning Research}, 18\penalty0
  (232):\penalty0 1--39, 2018.

\bibitem[Tibshirani(1996)]{lasso}
R.~Tibshirani.
\newblock Regression shrinkage and selection via the lasso.
\newblock \emph{Journal of the Royal Statistical Society: Series B
  (Methodological)}, 58\penalty0 (1):\penalty0 267--288, 1996.

\bibitem[Forgy(1965)]{forgy1965}
E.W. Forgy.
\newblock Cluster analysis of multivariate data: efficiency versus
  interpretability of classifications.
\newblock \emph{biometrics}, 21:\penalty0 768--769, 1965.

\bibitem[Huber(2011)]{huber2011robust}
P.J. Huber.
\newblock Robust statistics.
\newblock In \emph{International encyclopedia of statistical science}, pages
  1248--1251. Springer, 2011.

\bibitem[Fan et~al.(2017)Fan, Li, and Wang]{fan2017estimation}
J.~Fan, Q.~Li, and Y.~Wang.
\newblock Estimation of high dimensional mean regression in the absence of
  symmetry and light tail assumptions.
\newblock \emph{Journal of the Royal Statistical Society: Series B (Statistical
  Methodology)}, 79\penalty0 (1):\penalty0 247--265, 2017.

\bibitem[Loh(2017)]{loh2017statistical}
P.L. Loh.
\newblock Statistical consistency and asymptotic normality for high-dimensional
  robust {M}-estimators.
\newblock \emph{The Annals of Statistics}, 45\penalty0 (2):\penalty0 866--896,
  2017.

\bibitem[Sun et~al.(2019)Sun, Zhou, and Fan]{sun2019adaptive}
Q.~Sun, W.X. Zhou, and J.~Fan.
\newblock Adaptive huber regression.
\newblock \emph{Journal of the American Statistical Association}, 115\penalty0
  (2529):\penalty0 254--265, 2019.

\bibitem[Lozano et~al.(2016)Lozano, Meinshausen, and Yang]{Lozano2016}
A.C. Lozano, N.~Meinshausen, and E.~Yang.
\newblock {Minimum distance lasso for robust high-dimensional regression}.
\newblock \emph{Electronic Journal of Statistics}, 10\penalty0 (1):\penalty0
  1296 -- 1340, 2016.

\bibitem[Wang et~al.(2013)Wang, Wu, and Li]{Wang2013}
L.~Wang, Y.~Wu, and R.~Li.
\newblock {Robust variable selection with exponential squared loss}.
\newblock \emph{Journal of American Statistical Association}, 108\penalty0
  (502):\penalty0 632 -- 643, 2013.

\bibitem[Avella-Medina and Ronchetti(2018)]{AvellaMedina2018RobustAC}
M.~Avella-Medina and E.~Ronchetti.
\newblock Robust and consistent variable selection in high-dimensional
  generalized linear models.
\newblock \emph{Biometrika}, 105\penalty0 (1):\penalty0 31--44, 2018.

\bibitem[Prasad et~al.(2020)Prasad, Suggala, Balakrishnan, and
  Ravikumar]{Prasad2020RobustEV}
A.~Prasad, A.S. Suggala, S.~Balakrishnan, and P.~Ravikumar.
\newblock Robust estimation via robust gradient estimation.
\newblock \emph{Journal of the Royal Statistical Society: Series B (Statistical
  Methodology)}, 82\penalty0 (3):\penalty0 601 -- 627, 2020.

\bibitem[Belloni et~al.(2011)Belloni, Chernozhukov, and Wang]{bellobi2011}
A.~Belloni, V.~Chernozhukov, and L.~Wang.
\newblock {Square-root lasso: pivotal recovery of sparse signals via conic
  programming}.
\newblock \emph{Biometrika}, 98\penalty0 (4):\penalty0 791--806, 2011.

\bibitem[Bradic et~al.(2011)Bradic, Fan, and Wang]{bradic2011penalized}
J.~Bradic, J.~Fan, and W.~Wang.
\newblock Penalized composite quasi-likelihood for ultrahigh dimensional
  variable selection.
\newblock \emph{Journal of the Royal Statistical Society: Series B (Statistical
  Methodology)}, 73\penalty0 (3):\penalty0 325--349, 2011.

\bibitem[Wang et~al.(2012)Wang, Wu, and Li]{wang2012quantile}
L.~Wang, Y.~Wu, and R.~Li.
\newblock Quantile regression for analyzing heterogeneity in ultra-high
  dimension.
\newblock \emph{Journal of the American Statistical Association}, 107\penalty0
  (497):\penalty0 214--222, 2012.

\bibitem[Wang(2013)]{wang2013l1}
L.~Wang.
\newblock The {L}1 penalized {LAD} estimator for high dimensional linear
  regression.
\newblock \emph{Journal of Multivariate Analysis}, 120:\penalty0 135--151,
  2013.

\bibitem[Fan et~al.(2014)Fan, Fan, and Barut]{Fan2014ADAPTIVERV}
J.~Fan, Y.~Fan, and E.~Barut.
\newblock Adaptive robust variable selection.
\newblock \emph{Annals of statistics}, 42\penalty0 (1):\penalty0 324--351,
  2014.

\bibitem[Sun and Zhang(2012)]{sun2012scaled}
T.~Sun and C.H. Zhang.
\newblock Scaled sparse linear regression.
\newblock \emph{Biometrika}, 99\penalty0 (4):\penalty0 879--898, 2012.

\bibitem[Huber(1964)]{Huber1964RobustEO}
P.J. Huber.
\newblock Robust estimation of a location parameter.
\newblock \emph{Annals of Mathematical Statistics}, 35\penalty0 (1):\penalty0
  73--101, 1964.

\bibitem[Wang et~al.(2007)Wang, Li, and Jiang]{wang2007robust}
H.~Wang, G.~Li, and G.~Jiang.
\newblock Robust regression shrinkage and consistent variable selection through
  the {LAD}-lasso.
\newblock \emph{Journal of Business \& Economic Statistics}, 25\penalty0
  (3):\penalty0 347--355, 2007.

\bibitem[Zou and Yuan(2008)]{zou2008composite}
H.~Zou and M.~Yuan.
\newblock Composite quantile regression and the oracle model selection theory.
\newblock \emph{The Annals of Statistics}, 36\penalty0 (3):\penalty0
  1108--1126, 2008.

\bibitem[Jurečková(1971)]{jureckova1971nonparametric}
J.~Jurečková.
\newblock Nonparametric estimate of regression coefficients.
\newblock \emph{The Annals of Mathematical Statistics}, 42\penalty0
  (4):\penalty0 1328--1338, 1971.

\bibitem[Jaeckel(1972)]{jaeckel}
L.~A. Jaeckel.
\newblock Estimating regression coefficients by minimizing the dispersion of
  the residuals.
\newblock \emph{The Annals of Mathematical Statistics}, 43\penalty0
  (5):\penalty0 1449--1458, 1972.

\bibitem[Hettmansperger and McKean(1977)]{hettmansperger1977robust}
T.P. Hettmansperger and J.W. McKean.
\newblock A robust alternative based on ranks to least squares in analyzing
  linear models.
\newblock \emph{Technometrics}, 19\penalty0 (3):\penalty0 275--284, 1977.

\bibitem[Hettmansperger and McKean(1998)]{hettmansperge1998}
T.P. Hettmansperger and J.W. McKean.
\newblock \emph{Robust nonparametric statistical methods}.
\newblock Arnold, 1998.

\bibitem[Wang et~al.(2020)Wang, Peng, Bradic, Li, and Wu]{wlasso}
L.~Wang, B.~Peng, J.~Bradic, R.~Li, and Y.~Wu.
\newblock A tuning-free robust and efficient approach to high-dimensional
  regression.
\newblock \emph{Journal of the American Statistical Association}, 115\penalty0
  (532):\penalty0 1700--1714, 2020.

\bibitem[Wang and Li(2009)]{wwscad}
L.~Wang and R.~Li.
\newblock Weighted wilcoxon-type smoothly clipped absolute deviation method.
\newblock \emph{Biometrics}, 65\penalty0 (2):\penalty0 564--571, 2009.

\bibitem[Hunter and Lange(2004)]{MMtutorial}
D.R. Hunter and K.~Lange.
\newblock A tutorial on mm algorithms.
\newblock \emph{The American Statistician}, 58\penalty0 (1):\penalty0 30--37,
  2004.

\bibitem[Zhou(2010)]{multiWil}
W.~Zhou.
\newblock A multivariate wilcoxon regression estimate.
\newblock \emph{Journal of Nonparametric Statistics}, 22\penalty0 (7):\penalty0
  859--877, 2010.

\bibitem[Hettmansperger and McKean(1978)]{HM1978}
T.P. Hettmansperger and J.W. McKean.
\newblock Statistical inference based on ranks.
\newblock \emph{Psychometrika}, 43:\penalty0 69--79, 1978.

\bibitem[Hunter and Li(2005)]{HandL}
D.R. Hunter and R.~Li.
\newblock Variable selection using mm algorithm.
\newblock \emph{Annals of statistics}, 33\penalty0 (4):\penalty0 1617, 2005.

\bibitem[Yu et~al.(2015)Yu, Won, Lee, Lim, and Yoon]{MMlasso}
D.~Yu, J.H. Won, T.~Lee, J.~Lim, and S.~Yoon.
\newblock High-dimensional fused lasso regression using
  majorization-minimization and parallel processing.
\newblock \emph{Journal of Computational and Graphical Statistics}, 24\penalty0
  (1):\penalty0 121--153, 2015.

\bibitem[Chang et~al.(1999)Chang, McKean, Naranjo, and Sheather]{chang1999high}
W.H. Chang, J.W. McKean, J.D. Naranjo, and S.J. Sheather.
\newblock High-breakdown rank regression.
\newblock \emph{Journal of the American Statistical Association}, 94
  (445):\penalty0 205--219, 1999.

\bibitem[Heiler and Willers(1988)]{heiler1988}
S.~Heiler and R.~Willers.
\newblock Asymptotic normality of r-estimates in the linear model.
\newblock \emph{Statistics: A Journal of Theoretical and Applied Statistics},
  19 (2):\penalty0 173--184, 1988.

\bibitem[Johnson and Peng(2008)]{johnson2008rank}
B.A. Johnson and L.~Peng.
\newblock Rank-based variable selection.
\newblock \emph{Journal of Nonparametric Statistics}, 20 (3):\penalty0
  241--252, 2008.

\bibitem[Young(1981)]{alsc}
F.W. Young.
\newblock Quantitative analysis of qualitative data.
\newblock \emph{Psychometrika}, 46:\penalty0 357--388, 1981.

\bibitem[Zou and Hastie(2005)]{elastic}
H.~Zou and T.~Hastie.
\newblock Regularization and variable selection via the elastic net.
\newblock \emph{Journal of the Royal Statistical Society Series B (statistical
  methodology)}, 67\penalty0 (2):\penalty0 301--320, 2005.

\bibitem[Sherwood and Price(2020)]{MCENr}
B.~Sherwood and B.~Price.
\newblock \emph{mcen: Multivariate Cluster Elastic Net}, 2020.
\newblock URL \url{https://CRAN.R-project.org/package=mcen}.
\newblock R package version 1.2.

\bibitem[Friedman et~al.(2008)Friedman, Hastie, and Tibshirani]{glmnet}
J.~Friedman, T.~Hastie, and R.~Tibshirani.
\newblock Regularization paths for generalized linear models via coordinate
  descent.
\newblock \emph{Journal of statistical software}, 33\penalty0 (1):\penalty0 1,
  2008.

\bibitem[Li and Qin(2021)]{MTE}
S.~Li and Y.~Qin.
\newblock \emph{MTE: Maximum Tangent Likelihood Estimation for Linear
  Regression}, 2021.
\newblock URL \url{https://CRAN.R-project.org/package=MTE}.
\newblock R package version 1.0.1.

\bibitem[Chin et~al.(2007)Chin, Teschendorff, Marioni, Wang, Barbosa-Morais,
  Thorne, Costa, Pinder, Van~de Wiel, Green, et~al.]{chin07}
S.F. Chin, A.E. Teschendorff, J.C. Marioni, Y.~Wang, N.L. Barbosa-Morais, N.P.
  Thorne, J.L. Costa, S.E. Pinder, M.A. Van~de Wiel, A.R. Green, et~al.
\newblock High-resolution acgh and expression profiling identifies a novel
  genomic subtype of er negative breast cancer.
\newblock \emph{Genome biology}, 8\penalty0 (10):\penalty0 1--17, 2007.

\bibitem[Naranjo and Hettmansperger(1994)]{naranjo1994bounded}
J.D. Naranjo and T.P. Hettmansperger.
\newblock Bounded influence rank regression.
\newblock \emph{Journal of the Royal Statistical Society: Series B
  (Methodological)}, 56\penalty0 (1):\penalty0 209--220, 1994.

\bibitem[Sievers(1983)]{sievers1983weighted}
G.L. Sievers.
\newblock A weighted dispersion function for estimation in linear models.
\newblock \emph{Communications in Statistics-Theory and Methods}, 12\penalty0
  (10):\penalty0 1161--1179, 1983.

\bibitem[Chao et~al.(2021)Chao, Härdle, and Yuan]{Chao2021}
S.K. Chao, W.K. Härdle, and M.~Yuan.
\newblock Factorisable multitask quantile regression.
\newblock \emph{Econometric Theory}, 37(4):\penalty0 794--816, 2021.

\bibitem[Ding et~al.(2021)Ding, Qin, Wu, and Wu]{Ding2021}
H.~Ding, S.~Qin, Y.~Wu, and Y.~Wu.
\newblock Asymptotic properties on high-dimensional multivariate regression
  m-estimation.
\newblock \emph{Journal of Multivariate Analysis}, 183:\penalty0 104730, 2021.

\bibitem[Wang and Karunamuni(2022)]{Wang2022}
Y.~Wang and R.J. Karunamuni.
\newblock High-dimensional robust regression with lq-loss functions.
\newblock \emph{Computational Statistics and Data Analysis}, page 107567, 2022.

\bibitem[Tan et~al.(2022)Tan, Sun, and Witten]{Tan2022}
K.M. Tan, Q.~Sun, and D.~Witten.
\newblock Sparse reduced rank huber regression in high dimensions.
\newblock \emph{Journal of the American Statistical Association}, 118\penalty0
  (544):\penalty0 2383--2393, 2022.

\bibitem[Mishra and Müller(2022)]{Mishra2022}
A.~Mishra and C.L. Müller.
\newblock Robust regression with compositional covariates.
\newblock \emph{Computational Statistics and Data Analysis}, 165:\penalty0
  107315, 2022.

\bibitem[Zhao et~al.(2017)Zhao, Lian, and Ma]{Zhao2017}
W.~Zhao, H.~Lian, and S.~Ma.
\newblock Robust reduced-rank modeling via rank regression.
\newblock \emph{Journal of Statistical Planning and Inference}, 180:\penalty0
  1--12, 2017.

\bibitem[She and Chen(2017)]{She2017}
Y.~She and K.~Chen.
\newblock Robust reduced-rank regression.
\newblock \emph{Biometrika}, 104(3):\penalty0 633--647, 2017.

\end{thebibliography}






\end{document}